\documentclass[11pt]{aastex}   
\usepackage{graphicx,emulateapj5}
\usepackage{onecolfloat}

\shortauthors{McIntosh et al.}
\shorttitle{Evolution of Early-type Red Galaxies with GEMS}

\newcommand{\pegase}{{\sc P\'egase }}
\newcommand{\combo}{COMBO-17 }
\newcommand{\sersic}{{S\'{e}rsic }}
\newcommand{\hst}{{\it HST }}
\newcommand{\gim}{{\sc gim2d }}   
\newcommand\hkpc{$h^{-1}$\,kpc }   

\newcommand\ml{M$/$L }
\newcommand\mlns{M$/$L}

\newcommand\mlstar{M$_{\star}/$L }
\newcommand\mlstarns{M$_{\star}/$L}
\newcommand\lr{L$_V-r_{50}$ }
\newcommand\lrd{L$_V,r_{50}$ }
\newcommand\lrns{L$_V-r_{50}$}
\newcommand\mr{M$_{\star}-r_{50}$ }
\newcommand\mrd{M$_{\star},r_{50}$ }
\newcommand\mrns{M$_{\star}-r_{50}$}

\slugcomment{{\sc Draft: } \today }

\begin{document}


\def\head{

\title{The Evolution of Early-type Red Galaxies with the GEMS Survey: Luminosity--size and Stellar Mass--size Relations Since $z\sim1$}

\author{Daniel H.\ McIntosh$^1$, Eric F.\ Bell$^2$, Hans-Walter Rix$^2$,
Christian Wolf$^3$, Catherine Heymans$^2$, Chien Y.\ Peng$^{4,5}$, 
Rachel S. Somerville$^4$,
Marco Barden$^2$, Steven V.\ W.\ Beckwith$^{4,6}$, Andrea Borch$^2$,
John A.\ R.\ Caldwell$^{4,7}$, Boris H\"au{\ss}ler$^2$, Knud Jahnke$^9$,
Shardha Jogee$^{4,8}$, Klaus Meisenheimer$^2$,
Sebasti\'an F. S\'anchez$^{9,10}$, Lutz Wisotzki$^9$}
\affil{$^1$ Astronomy Department, University of Massachusetts, 
710 N. Pleasant St., Amherst, MA
01003, USA; \texttt{dmac@hamerkop.astro.umass.edu}}
\affil{$^2$ Max-Planck-Institut f\"ur Astronomie,
K\"onigstuhl 17, D-69117 Heidelberg, Germany\\
$^3$ Department of Physics, Denys Wilkinson Bldg., University
of Oxford, Keble Rd., Oxford, OX1 3RH, UK \\
$^4$ Space Telescope Science Institute, 3700 San Martin Drive, Baltimore MD, 21218, USA \\
$^5$ Steward Observatory, University of Arizona, 933 N. Cherry
Ave., Tucson AZ, 85721, USA \\
$^6$ Johns Hopkins University, Charles and 4th St., Baltimore, MD 21218, USA \\
$^7$ Present address: University of Texas, McDonald Observatory, Fort Davis, TX 79734, USA \\
$^8$ Present address: Department of Astronomy, University of Texas at Austin, 1 University Station C1400, Austin, TX 78712-0259, USA \\
$^9$ Astrophysikalisches Institut Potsdam, D-14482 Potsdam, Germany \\
$^{10}$ Present address: Centro Astronomico Hispano Aleman, E-04004 Almeria, Spain}

\begin{abstract}
We combine imaging from the {\it Hubble Space Telescope} Advanced Camera for
Surveys, as part of the GEMS (Galaxy Evolution from Morphologies and SEDs)
survey, with redshifts and rest-frame quantities from COMBO-17
to study the evolution of morphologically early-type galaxies 
with red colors since $z=1$. From $0.5\arcdeg \times 0.5\arcdeg$ imaging,
we draw a large sample of 728 galaxies with
centrally-concentrated radial profiles (i.e. $n\ge 2.5$ from \sersic fits) and
rest-frame $(U-V)$ colors on the red sequence.
We explore how the correlations of rest-frame $V$-band luminosity and of
stellar mass with intrinsic half-light size change over the last half
of cosmic time.
By appropriate comparison with the well-defined local relations from 
the {\it Sloan Digital Sky Survey}, we find that the {\it luminosity--size} and
{\it stellar mass--size} relations evolve in a manner
that is consistent with the passive aging of ancient
stellar populations. By itself, this result is consistent with 
a completely passive evolution of the red
early-type galaxy population. If instead, as demonstrated
by a number of recent surveys, the early-type galaxy population builds up
in mass by roughly a factor of two since $z \sim 1$, our results imply
that new additions to the early-type galaxy population follow similar
{\it luminosity--size} and {\it stellar mass--size} correlations, 
compared to the older subset of early-type galaxies. Adding early-type galaxies
to the red sequence through the fading of previously prominent disks
appears to be consistent with the data.  Through comparison with models, the
role of dissipationless merging is limited to $<1$
major merger on average since $z = 1$ for the most massive galaxies.
Predictions from models of gas-rich mergers are not yet mature
enough to allow a detailed comparison to our observations.
We find tentative evidence that the
amount of luminosity evolution depends on galaxy stellar mass,
such that the least massive galaxies show stronger luminosity
evolution compared to more massive early types. This could reflect a
different origin of low-mass early-type galaxies and/or younger stellar
populations; the present data is insufficient to discriminate between
these possibilities.
\end{abstract}

\keywords{galaxies: evolution --- galaxies: fundamental parameters (luminosities, stellar masses, radii) --- galaxies: general --- surveys}
} 

\twocolumn[\head]

\section{Introduction}

The formation and evolution of massive, early-type galaxies constitutes a
long-standing and crucial problem in cosmology. In all hierarchical models,
e.g. a $\Lambda$-dominated Cold Dark Matter ($\Lambda$CDM) cosmology, the
massive early-type galaxies seen now are expected to have formed through
mergers of smaller galaxies over time \citep{white91,cole00}. Yet, within these
models it is difficult to predict robustly at what point during this
assembly most stars formed. In order to constrain the star formation (SF) and
assembly histories of the early-type galaxy population, it is necessary to
explore the evolution of their number density and various scaling relations,
such as the fundamental plane
\citep{vandokkum96,kelson00c,vandokkum01a,treu02,wuyts04,vanderwel04}, the
{\it luminosity--size} relation
\citep{barrientos96,pahre96,schade97,barger98,schade99,ziegler99}, and the
{\it stellar mass--size} relation \citep{trujillo04a}. To this
end, we use the GEMS \citep[Galaxy Evolution from Morphology and
SEDs;][]{rix04} survey in conjunction with COMBO-17
\citep[Classifying Objects by Medium-Band Observations in 17 Filters;][]{wolf03,wolf04}
to construct the largest sample to date of distant $(0.2 < z \le 1.0)$
early-type galaxies with {\it Hubble Space Telescope} ({\it HST}) 
imaging to explore the evolution of
luminosity and stellar mass as functions of galaxy size.

Most initial studies of $z \le 1$ early-type galaxies focused on galaxies
in some of the richest clusters at each epoch
\citep[e.g.,][]{vandokkum96,vandokkum98b,kelson00c}, and found that the
mass-to-light ratios (\mlns) of massive galaxies in this environment
changed as expected for passive fading of a population formed exclusively
at very early epochs. In recent years, greatly expanded studies of ``red''
galaxies in random cosmological volumes have been carried out, extending
to redshifts of at least $z\sim 2$
\citep{chen03,rusin03,bell04b,drory04,fontana04,mccarthy04}. 
From colors and spectra,
these studies inferred that massive galaxies existed at all observed
epochs and that red galaxies at any given epoch contain almost exclusively
old (relative to the observational epoch) stellar populations. 
Yet, the age distribution of the stars in
these galaxies does not by itself speak to their dynamical assembly.

While the optical colors of massive red early-type galaxies would 
indeed be consistent with a population that
has evolved only through fading since z$\sim 2$, 
other data indicate that more must have happened.
In particular, the co-moving total stellar mass density in red-sequence galaxies
is lower at earlier epochs, amounting to a factor of $\sim 2$ buildup
since $z\sim 1$ \citep{chen03,bell04b,cross04}.
Several factors are expected to contribute to this evolution: 
(1) in galaxies with waning 
and disappearing SF, the disk will fade and the bulge
gain prominence in comparison; (2) spheroidal components
may newly form from gas-rich
disk mergers or from internal disk instabilities; 
or (3), galaxies within the red sequence
could undergo dissipationless merging,
changing their masses and structures while leaving their stellar
populations unchanged. All processes are expected to continue to the
present epoch in a hierarchical universe 
\citep[e.g.,][]{cole00,steinmetz02,khochfar03}.
The relative importance of the effects is expected to 
be a strong function of galaxy mass: gas-rich mergers, 
disk instabilities, and disk fading are expected to 
contribute strongly at low masses, whereas
dissipationless mergers may dominate
for high-mass galaxies \citep{khochfar03}.

The aging of a stellar population not only implies redder colors but also
dimming. We exploit this to apply an independent test for the population
evolution of massive galaxies. This test uses the morphological and
structural information for a large sample of early-type, red galaxies to
$z\sim 1$, which we draw from the COMBO-17 redshift and spectral energy
distribution (SED) survey 
\citep{wolf03,wolf04}
and the GEMS \hst imaging survey \citep{rix04}. By constructing the 
{\it luminosity--size} (\lrns) and {\it stellar mass--size} (\mrns)
relations for subsamples of morphologically selected early-type, red
galaxies at different redshifts, one can test the hypothesis of passive
evolution and provide constraints for other evolutionary scenarios. In
practice, the concentration of the light profile, specified by the \sersic
index $n\ge 2.5$ is used as a quantitative proxy for morphology.

For the most strict version of passive evolution (no new stars; no merging)
the prediction is clear: the \mr relation will remain unchanged
\footnote{This is true neglecting the possible change in effective radius
that might occur in a galaxy with a substantive age gradient. Stellar mass
loss in an aging stellar population would lead also to slow changes in
the \mr relation.} and the \lr relation should change in the sense that
galaxies of a given size would have been brighter by about 1 magnitude in
$V$-band at $z=1$, if the observed color evolution 
\citep{bell04b} is a guide to the stellar \ml
evolution. The predictions for the evolution of these relations in the
presence of merging are maturing rapidly.  Gas-free dissipationless
mergers \citep[see e.g.,][]{naab03,dantas03,nipoti03}
will slowly move galaxies away from the \lr and \mr 
relations \citep{nipoti03}; we discuss this issue 
in detail later.  Existing constraints
from the color-magnitude relation (CMR) or fundamental plane (FP)
limit the amount of dissipationless merging that can occur
to a factor of a few in terms of mass growth in this merging mode, 
if roughly equal-mass mergers dominate the mass growth
\citep{bower98,dantas03,nipoti03,gonzalez03,evstigneeva04}.
Predictions for gas-rich mergers
are less mature, and it is not clear what to expect for
the evolution of the \lr and \mr relations in this case.

As shown persuasively by \citet{simard99}, any such test for passive or
other physical evolution requires careful accounting for redshift-dependent
selection effects.
We will address this issue extensively, as the systematics of the sample
selection become increasingly important over diminishing statistical errors
for larger samples. Also, all existing surveys that are
deep enough to reach $z \sim 1$ cover such a small area that their volumes
at low redshifts ($z\le 0.2$) become too small. Therefore, we will combine
the COMBO-17/GEMS data at $z\ge 0.2$ with the present-day \lr
and stellar \mr relations for early-type galaxies \citep{shen03}
from the {\it Sloan Digital Sky Survey} \citep[SDSS;][]{york00}. Both for
\citet{shen03} and the $0.2<z<1.0$ sample presented here, stellar
mass-to-light ratios are estimated from the spectra and SEDs.

In this paper, we bring to bear the largest sample of morphologically and
color-selected early-type galaxies with \hst imaging for a comprehensive
analysis of the evolution of the \lr and \mr relations. Measuring the
evolution of relationships between these fundamental galaxy
properties over $0<z<1$ can place further constraints on galaxy formation
and evolution models. Specifically, we have two goals: (1) to quantify the
expected luminosity evolution of galaxies of a given size due to the simple
aging of their ancient stars with the best accounting of selection
effects to date; and (2) to test whether stellar mass evolution occurs at
different fixed galaxy sizes, which may provide constraints on the merging
history of these systems of old stars. The outline of this paper follows.
We describe our early-type galaxy data in \S2, including relevant
discussions of the GEMS imaging survey, the rest-frame quantities from
COMBO-17, the early-type sample selection, and our galaxy size
measurements. In \S3 we explore the evolution of the luminosity and
stellar mass scaling relations of early-type galaxies with $z\le1$. We
discuss our results in \S4 and give our conclusions in \S5. For rest-frame
luminosities, stellar masses, and physical sizes we assume a flat,
$\Lambda$-dominated cosmology with $\Omega_{\rm m} = 0.3$,
$\Omega_{\Lambda} = 0.7$, and $H_0 = 100 h $\,km\,s$^{-1}$\,Mpc$^{-1}$.
When estimating
expected passive stellar population evolution, we adopt a flat universe 
with the WMAP parameters 
\citep[$H_0 = 72$\,km\,s$^{-1}$\,Mpc$^{-1}$ and $\Omega_{\rm m} = 0.27$,][]{spergel03}, which give
a Hubble time of
13.5\,Gyr, to account for stellar \ml and color
evolution in the most realistic way possible at this time.

\section{Early-type Galaxy Data}
We select a well-defined sample of early-type galaxies with \hst imaging
from GEMS, and high-precision photometric redshifts and rest-frame
luminosities from COMBO-17.  In this
section we discuss briefly the high-resolution imaging and the aspects
of the \combo data relevant to this study.  Furthermore, we outline
the early-type sample selection and its completeness, and we describe
the galaxy size measurements from the \hst data.

\subsection{High-resolution Imaging}
\label{hstimaging}
The GEMS survey has obtained a large, two passband (F606W and F850LP) 
Advanced Camera for Surveys (ACS) image
mosaic over an area of $28\arcmin \times 28\arcmin$, encompassing the
Extended Chandra Deep Field South (E-CDFS).
This is the largest contiguous color map of the cosmos obtained with \hst and
consists of a grid of 78 mostly overlapping images taken
with the ACS wide field camera during November 2002.
The full details of the imaging grid, observations, data reduction,
calibration, and data quality assessment will be presented in 
Caldwell et al. (in preparation, C05).  An overview of the GEMS survey
experimental design is given in \citet{rix04}.  The foremost goal of GEMS is to
quantify the internal structural evolution of galaxies using statistically 
significant samples of galaxies at known redshifts.  
To this end we selected the E-CDFS field with an existing redshift data set
from COMBO-17. Furthermore, a
fraction of GEMS overlaps with the GOODS \citep{giavalisco04},
Ultra Deep Field (UDF), and UDF parallels \citep{bouwens04} 
programs, thus providing
a wealth of deeper, multi-wavelength data.

Hereafter, we concentrate our analysis on measuring galaxy sizes from the
longest rest-frame wavelength possible at which galaxy profiles 
appear most uniform 
and the contamination from star-forming regions in disk galaxies is minimized.
Therefore, we measure sizes by fitting simple models to the two-dimensional
galaxy imaging in the F850LP filter. 

For each pointing in the GEMS mosaic we have individual 
F850LP images, which are multi-drizzle combinations of three dithered
exposures (720--762 seconds each; see C05 for details). Briefly, the
exposures were processed individually prior to combining (i.e. bias and
dark current correction, flat fielding).
The final frames are free of most cosmic rays and remapped onto a
fine $0\farcs03$/pixel scale.  The full suite of 78 F850LP images are
quite flat with background differences between images considerably less 
than 0.2 ADU, and the local galaxy background measurements across each
frame have a typical RMS of 0.2 ADU.
In addition, we have a variance image (necessary for source detection)
with the same scale for each frame.  
The images are flux calibrated for photometric uniformity over the
large mosaic of ACS frames, which assures unbiased galaxy size measurements.
We calibrate the astrometry of each image to the
ground-based epoch J2000.0 system of COMBO-17, which allows
precise image/redshift cross correlation.  The reduced frames have
angular resolution of $0\farcs077$ FWHM, corresponding to a
physical scale of 700 pc 
at $z\sim0.75$, comparable to the resolution of Coma galaxy cluster
observations with $1\arcsec$ seeing.

The final GEMS source catalog contains
41,681 unique objects detected in the F850LP imaging.  
For source detection we employ SExtractor \citep{bertin96} in a two-step
strategy discussed in detail in \citet{rix04}.
Briefly, first we use a conservative (``cold'') detection and
deblending configuration that avoids spurious deblending of large galaxies
with strong features, yet is incomplete for faint, low surface brightness
objects found in \combo.  Next we use a ``hot'' configuration to detect
$99\%$ of known \combo sources down to a total apparent magnitude of
$R_{\rm tot}=24$ (Vega), the limit at which
the redshift performance drops dramatically.  The exact configuration
parameters are given in \citet{rix04}.  In each case, separate
catalogs for each GEMS pointing are constructed from the combined F850LP
image and a weight map ($\propto {\rm var}^{-1}$).  The use of weight maps
reduces the number of spurious detections in low signal-to-noise (S/N) areas 
of each image (e.g., near image edges). 

For each field we combine
all ``cold'' sources plus the subset of ``hot'' sources
residing outside of the isophotal area of any cold detections, and then we
remove duplicate sources to produce the final catalog with the
following properties: (1) contains 99\% of all
objects in the $R_{\rm tot}\le24$ mag \combo sample; (2) avoids
spurious deblending of large, bright galaxies exhibiting strong substructure
in \hst images from spiral features, etc.; and (3) provides a homogeneous,
flux and surface brightness limited catalog of all sources in the F850LP
ACS mosaic, regardless of \combo or other external information.
Through detailed simulations (H\"au{\ss}ler et al., in preparation,
hereafter H05) we find that
galaxy detection in GEMS is complete to effective surface brightnesses
in the F850LP-band of $\mu_{50}\sim 25$~mag~arcsec$^{-2}$ (within the
half-light radius) for $r^{1/4}$ spheroids and
$\mu_{50}\sim 24$~mag~arcsec$^{-2}$ for exponential disks.
The GEMS SExtractor source catalog will be published 
and described in complete detail in C05.

\subsection{Rest-frame Quantities from COMBO-17}
\label{c17data}
\combo is a deep photometric survey comprised of five (three completed)
disjoint fields, each $\sim0.25$ square degree in size, with flux measurements
in 5 broad and 12 medium passbands (between 3500 and 9000 angstroms).
This survey provides data for $\sim25,000$ galaxies with $R_{\rm tot}<24$ mag
in the range $0.2<z<1.2$.  The \combo data,
associated uncertainties, sample selection, and completeness are
discussed in detail in \citet{wolf04}.

The deep 17-passband data combined with stellar, AGN, and
non-evolving galaxy template spectra allow nearly all (98\%) objects to be 
assigned redshifts and SED classifications.  The redshift accuracies
depend primarily on $R$-band aperture magnitude, with
$\delta z/(1+z)\sim 0.02$ for $R_{\rm ap}<22$ mag. When going fainter the
errors reach $\delta z/(1+z)\sim 0.05$ at $R_{\rm ap}=24$ mag \citep{wolf04}.
Reliable redshift estimates allow the construction
of rest-frame luminosities and colors.  Depending on the redshift in
question, rest-frame luminosities are determined either by interpolation
or mild extrapolation.  For this study we
use the rest-frame $V$-band absolute magnitudes for the luminosity
parameter, which are extrapolated for galaxies with $z>0.7$.
Nevertheless, the leverage of the \combo filter set continues to provide
accurate $V$-band magnitudes to $z=1$, with typical uncertainties of
$\sim10\%$ ($z<0.7$) and $\sim15\%$ ($z>0.7$).  We
note that the quoted redshift accuracies correspond to $\la15\%$ fractional
distance errors for $z>0.2$.

We cross correlate the positions of sources from the GEMS F850LP imaging
with source coordinates from the \combo E-CDFS catalog\footnote{Available at
http://www.mpia-hd.mpg.de/COMBO/combo\_index.html.}.
We find 6152 galaxies with redshifts $0.2\leq z \leq 1.0$ and 
$R_{\rm ap}\leq24$ mag.
This is our overall galaxy sample.  We note that the \combo magnitude
limit corresponds to the reliability of estimating redshifts, which
is brighter than the limit to detect galaxies with high
completeness both in \combo and GEMS.  For red-selected galaxies,
which are the primary concern of this paper, the completeness limit owing to
redshift reliability depends somewhat on redshift: for $0.2<z<0.6$ the mean
90\% limit occurs at $R_{\rm ap}\sim23.5$, while at larger redshifts 
($0.6<z<1.0$) the sample is fully complete down to $R_{\rm ap}=24$ mag.
At redshifts above $z=1$, \combo remains complete to the
magnitude limit, yet this limit begins restricting the sample to only 
the very luminous ($M_V<-21+5\log_{10}h$ at $z\sim1$) red galaxies.
We therefore adopt a strict $z=1$ cut for our analysis.

We use stellar mass M$_{\star}$ estimates based on direct modeling
of the \combo SEDs (Borch et al., in preparation).
The 17-passband \combo SEDs were compared with a library of galaxy template
spectra obtained from the \pegase code \citep[see][for a description of an 
earlier version of the code]{fioc97}. From each galaxy's SED we estimate
directly its stellar \ml ratio (\mlstarns), and hence M$_{\star}$; M$_{\star}$ 
uncertainties include contributions from redshift uncertainty. We use 
a \citet{kroupa93} stellar initial mass function (IMF), which produces
\mlstar values comparable to those from a \citet{kroupa01} IMF, and
$\sim 0.3$ dex lower than from a \citet{salpeter55} IMF.
For red-selected galaxies, these stellar masses agree to within $\la 40$\%
(0.11 dex), and show no average offset when compared to stellar masses based
on the simple relation between $(B-V)$ color and \mlstar from \citet{bell03b},
assuming the same IMF.
The stellar mass estimates suffer from a number of random and systematic
uncertainties, such as uncertainties in galaxy age, bursts of SF
in the last 1-2 Gyr, and variations in the
relationship between dust reddening and extinction 
\citep[see e.g.,][for a more in-depth discussion of the relevant 
sources of uncertainties]{bell01,bell03b}. Generally these uncertainties amount
to $\sim 0.3$ dex; for the red-sequence galaxies studied here, it is
likely that the stellar mass estimates are more accurate than
this, as relatively old stellar populations are somewhat easier to model
robustly with simplistic SFH prescriptions.

\subsection{Early-type Galaxy Selection}
In very generic terms, the visual appearances of galaxies fall into two broad 
types: (1) early-type refers to a galaxy with spheroid or
bulge-dominated (centrally-concentrated) morphology; and (2) late-type
corresponds to a disk-dominated or irregular system.  In the local universe,
several classes from the familiar \citet{hubble26} sequence are considered 
early-type in morphology -- ellipticals (E), lenticulars (S0),
and spirals with dominant bulges and small disks with tightly wound spiral
arms (Sa).  In addition to physical appearance, the optical colors of
galaxies are broadly different -- early types have typically
red colors suggestive of a dominant population of old stars with little or
no current SF; late types tend to be bluer in color as a result of
ongoing SF and some fraction of young stars.  This
bimodal distribution of galaxies in color space is observed both
locally \citep{strateva01,hogg02,blanton03d,baldry04} and out to $z\sim1$
\citep{bell04b,weiner04}. Furthermore, red galaxies at $z\sim0.7$
contain the same predominance ($\sim85\%$ by optical luminosity density) 
of early-type galaxies (E/S0/Sa) as locally \citep{bell04a}, with a
small contamination by edge-on disk galaxies with colors reddened by internal
extinction.  For this
study we are interested in early-type galaxies with red colors, thus we
select our sample using both color and morphology.

\subsubsection{Red Sequence}
Red-color selection permits an objective and empirical first cut for defining
early-type galaxies over the last half of cosmic history.  
We use the empirical fit to the CMR evolution
(including local comparison points) from \citet{bell04b}\footnote{
This cut is slightly bluer than that given by \citet{bell04b}. Owing partially
to uncertainties in photometric calibration and galactic
foreground extinction, the red sequence has slightly different
colors in the three COMBO-17 fields, and the cut we use here is
more appropriate for the E-CDFS.}
to select red-sequence galaxies at a given redshift $z$ 
with rest-frame colors redder than
\begin{equation}
(U-V) = 1.05 - 0.31z -0.08(M_V - 5\log_{10}h + 20) .
\label{colorcut}
\end{equation}
This color cut is 0.25 mag redder than the fit to the CMR evolution, and
similar in philosophy to the \citet{bo84} criteria, except that we select
red rather than blue galaxies.
The color evolution of red-sequence galaxies selected with this
definition is in good agreement with the expected evolution of a single-age
stellar population anchored to $z=0$. There are 1166 red-sequence 
objects with $0.2<z\le1.0$ and GEMS imaging.

To facilitate meaningful profile fitting,
we visually inspect the F850LP images of the red-sequence sample.
\combo is expected to misclassify a small number of red M-class stars as
moderate-redshift red-sequence galaxies \citep{wolf04}. With the superior
resolution of GEMS, we can directly detect 69 misclassified stars, which
are shown in Figure \ref{Obsvd} for completeness, and are omitted from the
analysis hereafter. We note that these stars are also separated cleanly from
galaxies using a SExtractor CLASS\_STAR$\ge0.85$ cut. 
Additionally, we remove 50 red-sequence galaxies with poor S/N found 
near GEMS image edges
or within inter-chip gaps. This leaves 1047 red-sequence galaxies with
useful GEMS imaging. Last,
we remove 84 galaxies with morphologies where reliable fits are
not possible (14 with prominent dust lanes, 29 irregulars,
and 41 peculiar/interacting).  We are left with a final sample of 963
morphologically ``normal'' red galaxies with the following breakdown of coarse
types: 699 E/S0, 162 Sa, and 102 late-type spirals (Sb--Sd).  Owing to
the subjective nature of visual classifications, we divide the sample
morphologically using a quantitative method in the next section.

\subsubsection{Quantitative Morphology}
\label{QMorph}
We refine our sample further by determining which red-sequence galaxies
have early-type morphology using a quantifiable and repeatable method.
The simple \citet{sersic68} model, which describes the radial surface 
brightness profile of a galaxy by $\Sigma (r) \propto \exp{(-r^{1/n})}$, is
the most widely-used parametric function.
The \sersic index $n$ describes
how centrally-concentrated a galaxy appears, with $n=4$ describing the
familiar $r^{1/4}$ profile of ellipticals \citep{devauc48}, and $n=1$
representing the exponential profile commonly seen in spiral disks.  
Studies with the SDSS
have adopted empirical cutoffs around $n=2.5$ to separate morphologically
early and late types \citep{blanton03d,shen03,hogg04}.  In addition,
using a sample of nearly 1500 GEMS galaxies in the thin redshift slice 
$0.65\le z\le 0.75$, \citet{bell04a} showed that an $n=2.5$ cut was reliable
at distinguishing between visually-classified early and late types.
To remain consistent, we
adopt $n\ge 2.5$ to denote early types in this study.

We note that the $n=2.5$ cut is a crude concentration cut that takes no
account of the real diversity of galaxy forms seen in nature.  In particular,
this cut does not account explicitly for dynamically-distinct bulge and disk 
components, and we reserve a detailed fitting of the photo-bulge and
photo-disk components to future papers.
Nevertheless, it is unclear
whether bulge/disk decompositions will be any more robust or physical
than the simple $n=2.5$ cut.  We note two additional sources of concern with
the $n\ge2.5$ selection.  First, $(1+z)^4$ dimming of disks may cause
a systematic bias such that large disks at high redshift may be too
faint to detect and, thus, appear as bulge-dominated galaxies (larger $n$ 
values).  Second, a correlation exists
between $n$ and absolute $B$-band magnitude for visually-selected 
ellipticals \citep[e.g.,][]{trujillo04b}.  Therefore, a strict $n=2.5$ cut
is effectively a luminosity cut of roughly $M_V=-17+5\log_{10}h$.  This 
magnitude limit is below our redshift reliability cutoff at all redshifts
except for the $z\le0.3$ interval, at which the
drop in numbers of galaxies fainter than $-17.5+5\log_{10}h$ may be
explained by this effect.
Ultimately, we accept these caveats with the understanding
that we have a repeatable and empirically-motivated early-type galaxy sample.

In Appendix \ref{fitting}, we describe our method for
fitting a \sersic model to each galaxy's two-dimensional luminosity
profile from the F850LP imaging.  Briefly, we use the fitting code
\gim \citep{simard02} to fit the sample of 963 red-sequence galaxies and
find that 96\% (928) are fit successfully.  The majority of fitting failures
are the result of \gim reaching the \sersic parameter limit of $n=8$.
We estimate uncertainties of $\delta n/n \sim 0.25$. Our final
sample contains 728 red galaxies out to $z=1$ with $n\ge 2.5$.
In general, using an $n=2.5$ cut does a good job
separating red-selected galaxies into morphologically early and late types.
We find only 3\% of the red $n\ge 2.5$ sample have late-type (Sb-Sd)
visual morphologies, while 14\% of the visually early types (E-Sa)
have $n<2.5$; this is typical of what was found in the comparison of
by-eye and \sersic classifications at $z\sim0.7$ by \citet{bell04a}.
We show in Figures \ref{Sthumbs} -- \ref{Lthumbs} that representative
examples of the red GEMS galaxies with $n\ge 2.5$ are visually early-type
in appearance. For contrast, in Figure \ref{Dthumbs} we display a
handful of the red-sequence galaxies with $n<2.5$ that correspond
to edge-on disks.

\subsection{Effective Size Measurements}
\label{sizes}

\subsubsection{Galaxy Sizes}
The main focus of this work is the measurement and analysis of galaxy
sizes as a function of redshift.
\gim calculates the half-light semi-major axis length $a_{50}$
by integrating the best-fit \sersic model flux to infinity.  
For meaningful comparison
with SDSS results we adopt a circular half-light radius, or geometric mean,
given by $r_{50}=a_{50}\sqrt{1-e}$, where $e$ is the best-fit model 
ellipticity.
We have applied a robust method for determining the systematic
and random uncertainties of each model parameter by measuring their random
offset and variance with respect to known values using a large sample
of artificial galaxies analyzed in the same manner as the GEMS data (H05).
This method has proven successful in other galaxy fitting
work \citep[e.g.,][]{tran03,macarthur03}.  We find that $r_{50}$ has typical
random uncertainties of $\sim35\%$ with no systematic error.  In addition,
we find that \gim $r_{50}$ measurements are reliable for $>90\%$ of galaxies
with F850LP-band surface brightness of $\mu_{50}\le 24$ mag asec$^{-2}$.
As seen in Figure \ref{Obsvd}, our early-type sample is well within this limit.

Sky level uncertainties are an important source of systematic error in size
measurements when applying profile fitting techniques 
\citep[e.g.,][]{dejong96c}. We estimate the error in our fitting-derived sizes
due to the uncertainty in our local background measurements
($\sigma_{\rm bkg}\sim0.2$ ADU) by repeating the fits with the sky level
held to constant values of $\pm1\sigma_{\rm bkg}$. We find that our size
measurements have average uncertainties due to the sky of $\sim10\%$.

Finally, our galaxy sizes are based on $r_{50}$ in a fixed, observed
passband (F850LP), which corresponds roughly to the SDSS $z$-band.  
Therefore, galaxy sizes at different redshifts
are measured at different rest-frame wavelengths; e.g., the
central wavelength of the F850LP filter corresponds to rest-frame 7560~\AA
at $z=0.25$ and 4846~\AA at $z=0.95$. For a consistent
comparison with the local scaling relations from \citet{shen03}, we correct
each $r_{50}$ measurement to rest-frame $r$-band. Early-type galaxies are
known to have internal radial color gradients 
\citep[e.g.,][]{franx89,peletier90,bernardi03d}, 
with redder colors towards their centers.
Such gradients should result in a wavelength-dependent galaxy size.
We estimate a correction to shift $z$-band sizes to rest-frame $r$-band
based on typical color gradients of
$\Delta (B-R)/\Delta \log{r} = -0.09$ and $\Delta (U-R)/\Delta \log{r} = -0.22$
from \citet{peletier90}.  
We assume an $r^{1/4}$ surface brightness profile and
find that this gradient will produce measured sizes in the $U$-band to be
about 6.5\% bigger than in the $r$-band; i.e. 
$\Delta r/\Delta \log{\lambda} = -0.25$.  Between $z=1$ and $z=0$ the
same observed passband corresponds to a factor of 2 change in wavelength, or
a 7.5\% size correction per unit redshift.  The correction to rest-frame
$r$-band ($z$-band observed at $z\sim0.45$) is then given by
\begin{equation}
\frac{\Delta r}{r_{\rm obs}} = -0.075(z-0.466) .
\end{equation}
Owing to the mild color gradients this correction is quite small,
amounting to only 4.0\% (1.5\%) decrease (increase) in observed
size for $z=1$ ($z=0.25$), and as such it does not significantly affect our 
conclusions.  We note that ellipticals with blue cores 
(i.e. inverted color gradients) have been observed by \citet{menanteau01b} 
in the Hubble Deep Field (HDF), yet these galaxies are
globally bluer than our red-sequence cut.  The assumptions we make regarding
color gradients are based on red galaxies; therefore, it is fair to 
apply this correction to our red-selected sample.

\subsubsection{Magnitude-Size Distributions}
Before proceeding with our analysis of scaling relations using rest-frame
quantities, we plot in Figure \ref{Obsvd} the
magnitude-size distributions in the observed F850LP-band frame
for all 928 red-sequence galaxies (split into eight bins of width $\Delta z=0.1$
spanning redshifts $0.2<z\le1.0$) to illustrate several aspects of
our sample.  We calculate the apparent magnitude $m_{850}$ of each galaxy from
the total intensity of the best-fit model (see Appendix \ref{fitting}).
We see that galaxy apparent magnitudes and angular sizes are correlated out
to $z=1$, such that bigger galaxies are brighter.  We note that
the effects of $(1+z)^4$ cosmological dimming appear as a shift of the
magnitude-size correlation towards lower surface brightnesses at increasing
redshifts. Specifically, between $z=0.25$ and $z=0.75$ the mean relation has
faded by the expected 1.5 mag. Furthermore, nearly all
galaxies have $r_{50}<1\farcs$0, clearly illustrating the need for \hst
resolution to obtain accurate size measurements for distant galaxies.  All
red-selected galaxies are well-resolved, and have effective surface 
brightnesses $>1.0$ mag brighter
than the \gim reliability limit and $>2.0$ mag brighter than the 85\%
detection completeness limit.  Therefore, our early-type sample is limited
only by the $R_{\rm ap}=24$ mag cutoff (corresponding to $m_{850}\sim22$) for
reliable photometric redshifts inherent to the \combo survey.

In addition, we show where $n<2.5$ (open squares in 
Figure \ref{Obsvd}) red-selected galaxies are
found in the magnitude-size plane.  Most red $n<2.5$ galaxies tend to
have lower surface brightnesses relative to the spheroid-dominated
$n\ge 2.5$ galaxies (solid symbols), and are for
the most part highly-inclined disk galaxies. 
These results are in qualitative agreement with the different magnitude-size
correlations for reddened edge-on disks and early-type galaxies in the
local universe \citep{blanton03d}.

\begin{figure*}
\center{\includegraphics[scale=0.8, angle=0]{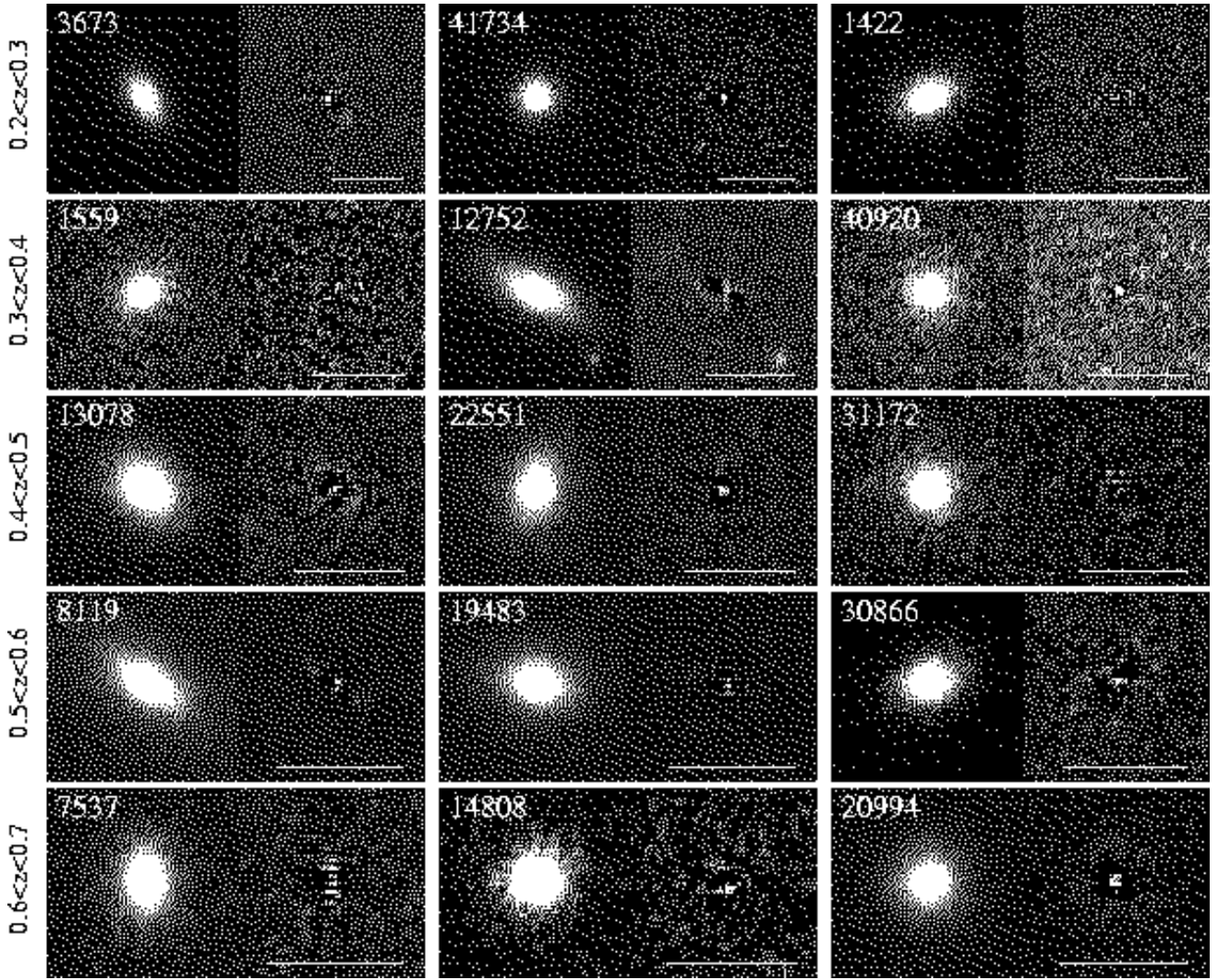}}
\caption[]{Representative examples of small 
(fixed-size bin {\sc i}: $0.5<r_{50}\le1.0\ h^{-1}$\,kpc)
early-type ($n\ge 2.5$ profiles with red-sequence colors)
galaxies in the redshift range $0.2< z \le 0.7$.  All galaxies shown
are visually early types.  In each row we
show the F850LP-band galaxy and \gim fitting residual postage stamp images for
three randomly-selected galaxies in each redshift interval.
We give the galaxy \combo identification number for each galaxy in the upper
left.
The postage stamp images are $7 h^{-1}$\,kpc per side.
The horizontal white lines represent 1 arcsecond.
\label{Sthumbs}}
\vspace{-0.2cm}
\end{figure*}

\begin{figure*}
\center{\includegraphics[scale=0.8, angle=0]{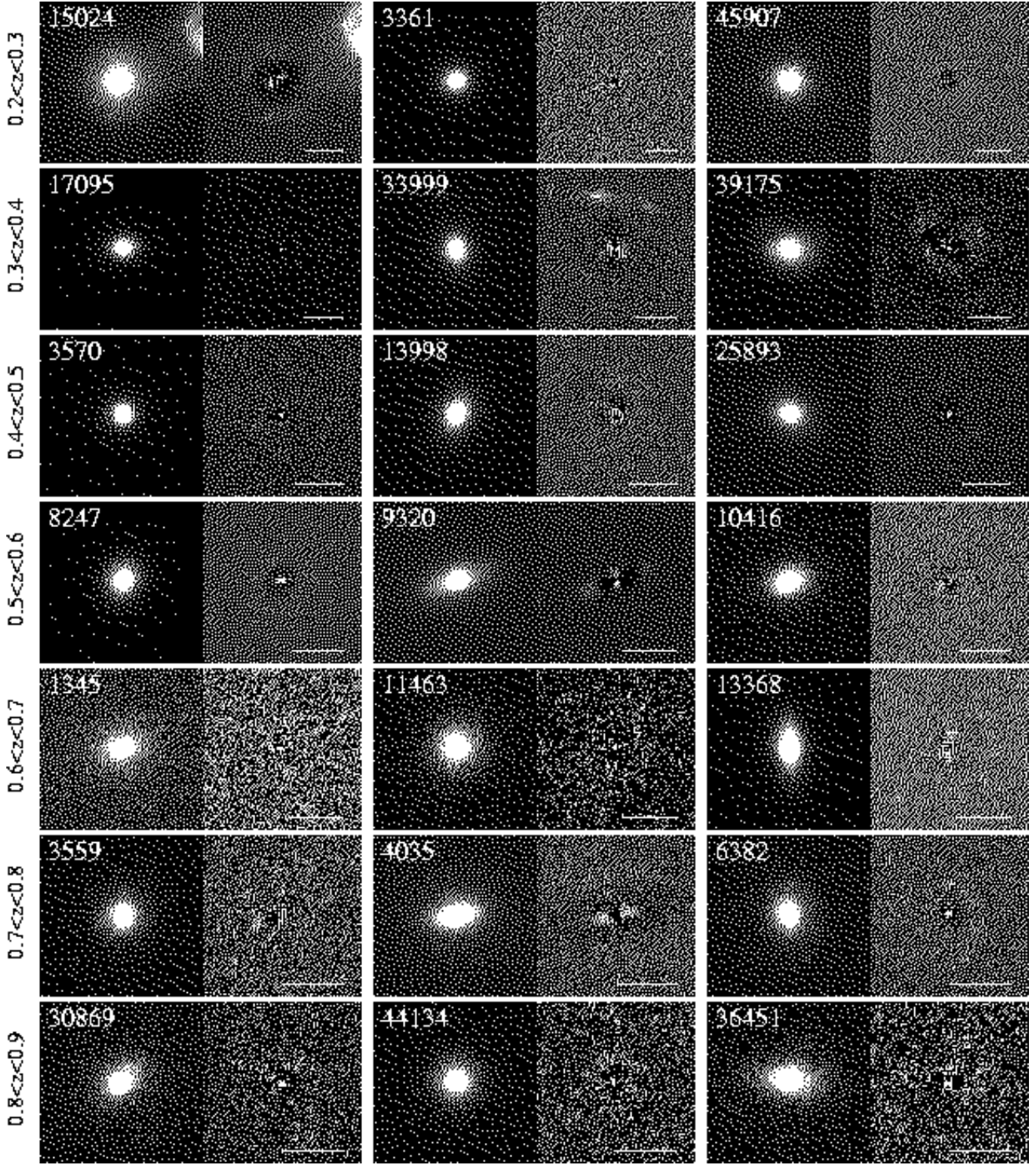}}
\caption[]{Representative examples of medium size
(bin {\sc ii}: $1.0<r_{50}\le2.0\ h^{-1}$\,kpc)
early-type galaxies with $n\ge 2.5$ profiles and red colors
spanning redshifts $0.2< z \le 0.9$. All galaxies shown
are visually early types.  Three example images with their
corresponding fitting residuals
are shown from each redshift interval as in Figure \ref{Sthumbs}.
The postage stamp images are $14 h^{-1}$\,kpc per side, with 1 arcsecond
denoted by horizontal white lines.
\label{Mthumbs}}
\vspace{-0.2cm}
\end{figure*}

\begin{figure*}
\center{\includegraphics[scale=0.8, angle=0]{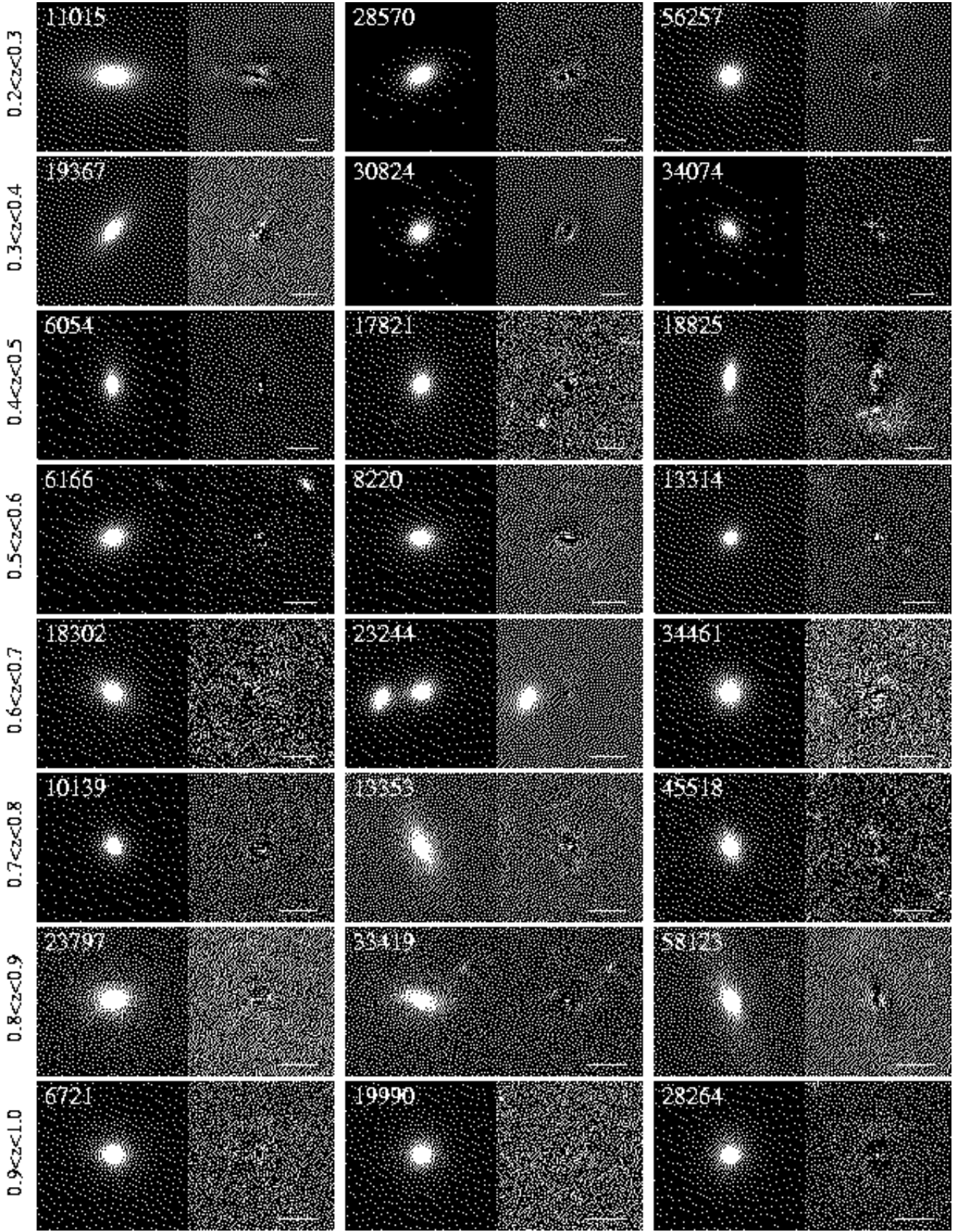}}
\caption[]{Representative examples of large size
(bin {\sc iii}: $2.0<r_{50}\le5.0\ h^{-1}$\,kpc)
early-type galaxies with $n\ge 2.5$ profiles and red colors
spanning redshifts $0.2< z \le 1.0$. All galaxies shown
are visually early types.  Galaxy and fitting residual images are
shown from each redshift interval as in Figures \ref{Sthumbs} and \ref{Mthumbs}.
The postage stamp images are $21 h^{-1}$\,kpc per side, with 1 arcsecond
denoted by horizontal white lines.
\label{Lthumbs}}
\vspace{-0.2cm}
\end{figure*}

\begin{figure*}
\center{\includegraphics[scale=0.8, angle=0]{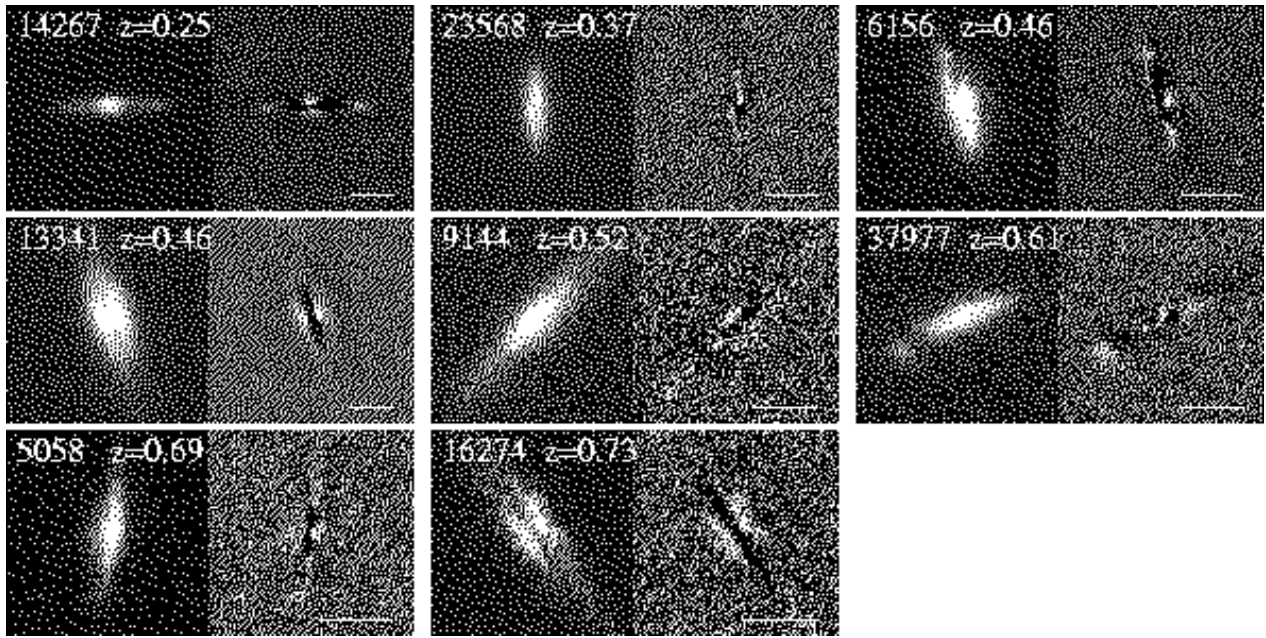}}
\caption[]{Examples of late-type disk galaxies ($n<2.5$ profiles)
with red-sequence colors.  We
show the F850LP-band galaxy and \gim fitting residual postage stamp images 
for 8 edge-on disks spanning redshifts $0.25\le z\le 0.73$.
We give the galaxy \combo identification number and redshift for each galaxy.
The postage stamp images are $14 h^{-1}$\,kpc per side.
The horizontal white lines represent 1 arcsecond.
\label{Dthumbs}}
\vspace{-0.2cm}
\end{figure*}

\begin{figure*}
\center{\includegraphics[scale=0.8, angle=0]{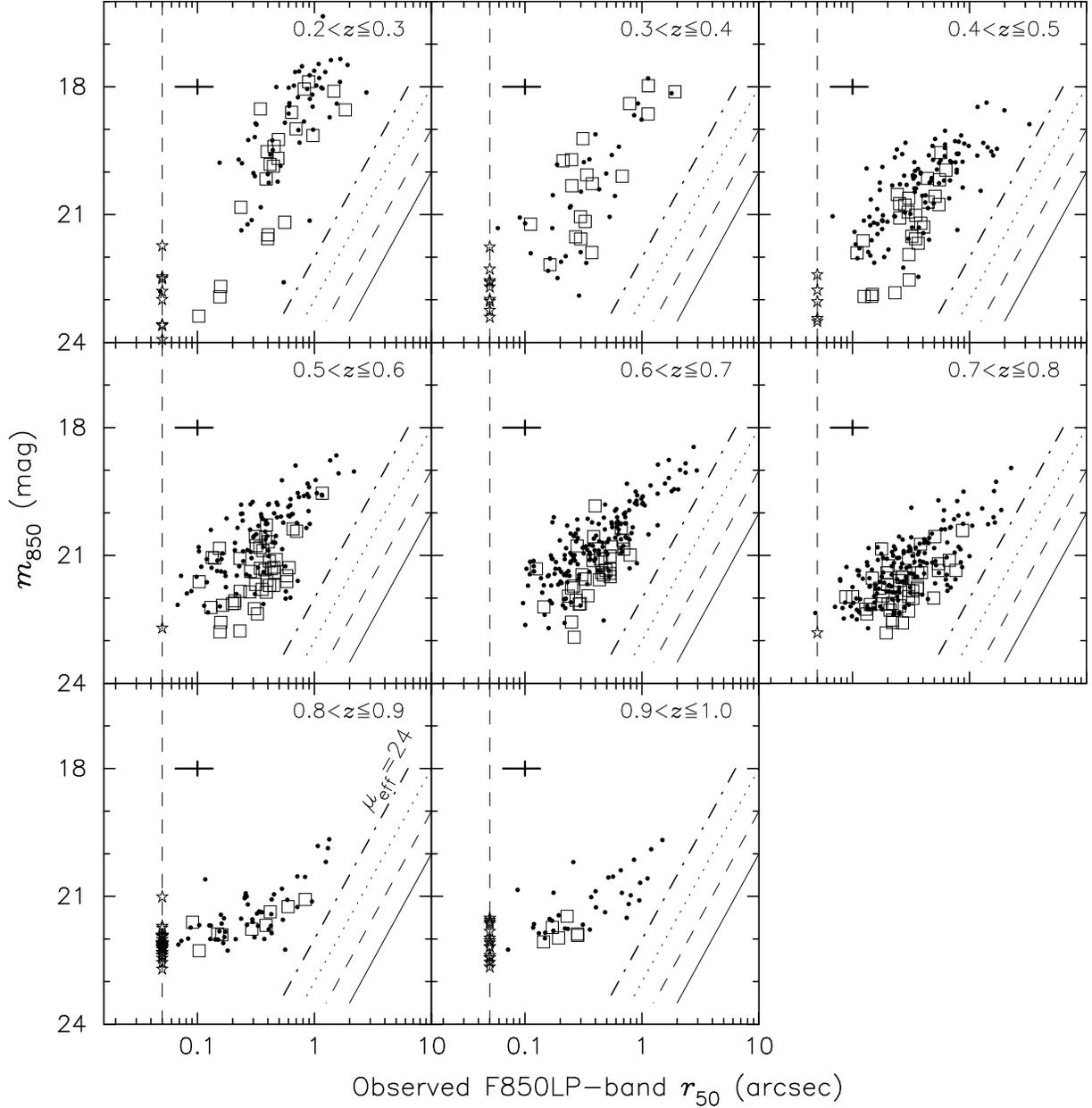}}
\caption[]{Observed magnitude-size distributions for 928 red-sequence galaxies
with robust \gim fits spanning redshifts $0.2<z\le1.0$ split into eight 
$\Delta z=0.1$ bins.
In each panel we plot the effective size $r_{50}$ 
(geometric mean half-light) and apparent magnitude $m_{850}$ from \sersic profile
fits to the F850LP-band ACS imaging data. We show the expected uncertainties
of the fit quantities with error bars at the upper left of each panel.
We denote $n\ge 2.5$ galaxies
with solid points and those with $n<2.5$ profiles with open squares.
The PSF radius ($0\farcs05$) is given by the dashed
vertical line.  Through simulations (see H05) we have established a
surface brightness limit of $\mu_{50}=24$ mag asec$^{-2}$ (diagonal dot-dashed
line) for reliable fits with {\sc gim2d}.  The additional diagonal
lines show surface brightnesses of $\mu_{50}=25$ (dotted),
$\mu_{50}=26$ (dashed), and $\mu_{50}=27$ mag asec$^{-2}$ (solid), which
represent the 85\%, 60\%, and 0\% detection completeness limits, respectively,
from simulations of $n=4$ profiles \citep[see][]{rix04}.  
We conclude that our early-type sample
selection is not surface-brightness limited. For completeness, we include the 
69 stars misclassified as red-sequence galaxies by COMBO-17 (open stars);
we remove these from our further analysis.
\label{Obsvd}}
\vspace{-0.2cm}
\end{figure*}

\section{Scaling Relations of Early-type Galaxies With $0.2<z\le1.0$}
\label{Analysis}

\subsection{Luminosity--Size Relation}
\label{LrRel}
\begin{figure*}
\center{\includegraphics[scale=0.75, angle=0]{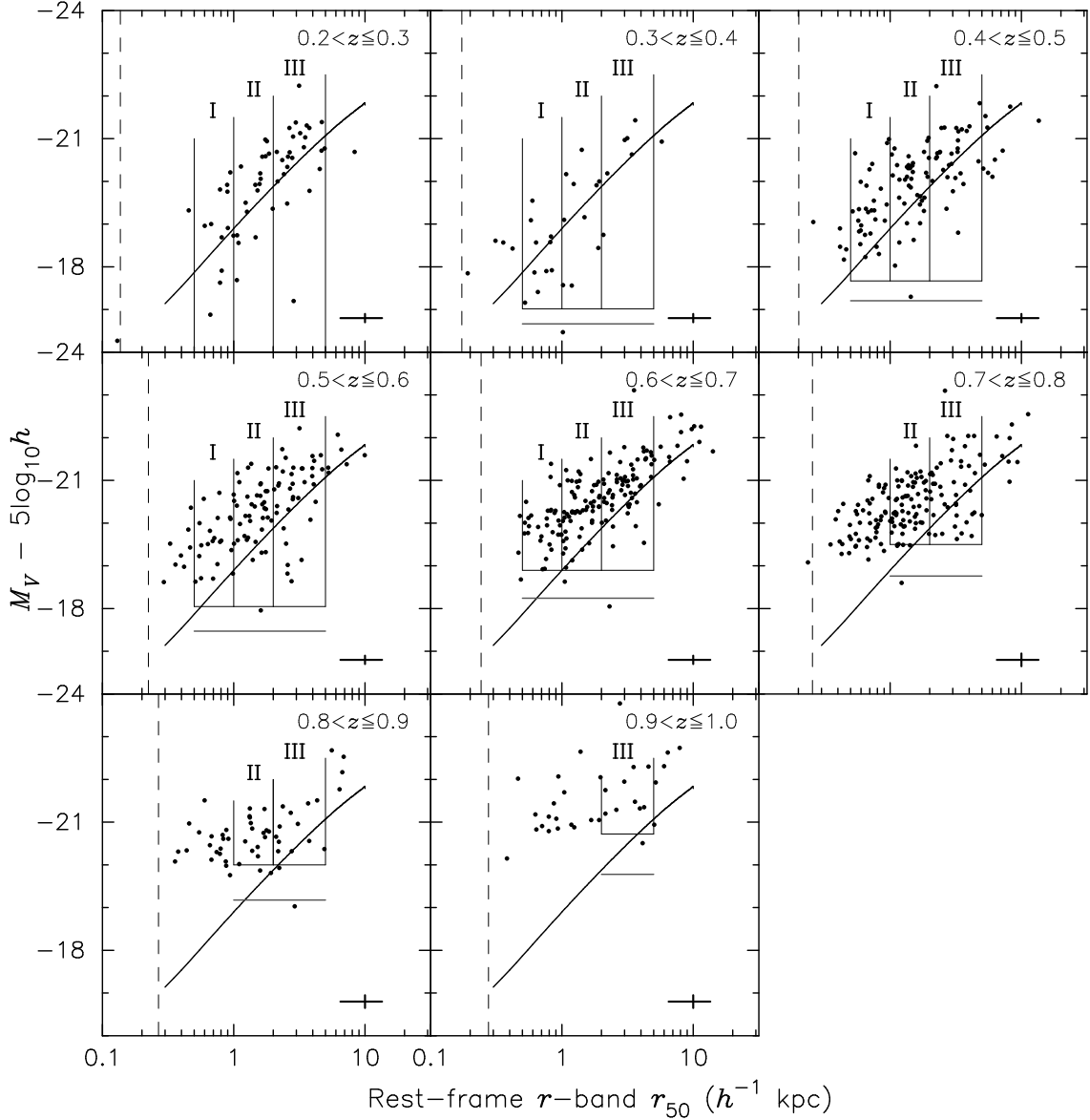}}
\caption[]{Rest-frame {\it luminosity--size} relations for 
728 early-type galaxies
in eight redshift epochs spanning $0.2<z\le1.0$.  All galaxies have $n\ge 2.5$
profiles and red-sequence colors.  In each panel we plot the physical
size, from \gim fits to F850LP-band images (corrected to rest-frame $r$-band),
against $V$-band absolute magnitude.
The error bars shown at the bottom right of each panel
give the 35\% $r_{50}$ uncertainty, and the 10\%
(interpolated for $z<0.7$) or 15\% (extrapolated
for $z>0.7$) $M_V-5\log_{10}h$ uncertainties.
The ridgeline for the $z\sim0$ \lr
relation for early types ($n\ge2.5$) from \citet{shen03} is shown in each panel
for comparison (diagonal line). 
The present-day ridgeline is the median $M_V-5\log_{10}h$ at each $r_{50}$ from
our mock catalog (see left panel of Figure \ref{Mock}). The vertical
lines delineate the fixed-size bins: ({\sc i}) $0.5<r_{50}\le1.0\ h^{-1}$\,kpc;
({\sc ii}) $1.0<r_{50}\le2.0\ h^{-1}$\,kpc; and ({\sc iii}) $2.0<r_{50}\le5.0\
h^{-1}$\,kpc.  The horizontal line attached to the vertical lines 
gives the faint limit of each
redshift-dependent completeness map, corresponding to the $R_{\rm ap}=24$ mag
\combo redshift reliability limit at each redshift interval.  Therefore, the
open-ended rectangles outline the non-zero completeness regions of
L$_V,r_{50},z$-space, and the additional isolated horizontal line represents 
the faint limit
of the completeness map passively faded to $z=0$. We show the physical size
corresponding to the PSF radius ($0\farcs05$) at each redshift (vertical
dashed line).
\label{Lr8}}
\vspace{-0.2cm}
\end{figure*}

We now present the \lr relations for early-type galaxies with
$0.2<z\le1.0$ and we will compare these to the present-day distribution.
In Figure \ref{Lr8}, we divide our sample into eight redshift bins of 
$\Delta z=0.1$ and for each we
plot the absolute $V$-band magnitude against the rest-frame $r$-band
size in physical units ($h^{-1}$\,kpc).
The bulk of the early-type population at each 
epoch spans roughly a 0.5 to 5.0 \hkpc half-light size distribution, 
with a handful of smaller ($<0.5 h^{-1}$\,kpc) and larger ($>5.0 h^{-1}$\,kpc)
galaxies in some redshift slices. Starting at low redshift, the average number
of galaxies per redshift slice increases as we sample larger co-moving
volumes, until $z\sim0.75$ where the counts start falling
off as expected in a magnitude-limited sample.  The second lowest redshift
bin is quite sparsely-populated as a result of large-scale fluctuations.

For comparison, in each redshift interval of Figure \ref{Lr8} we show the 
``ridgeline'' of the present-day \lr relation for early-type galaxies.
The $z\sim0$ ridgeline is given by the median 
absolute $V$-band magnitude as a function of galaxy size using a mock
catalog of $M_V-5\log_{10}h$ and $r_{50}$ values that follow the 
luminosity function of SDSS-selected, red-sequence galaxies
from \citet{bell04b}, and the \lr relation for
$n\ge2.5$ galaxies in SDSS from \citet{shen03}.
Using a complete sample of $n\ge2.5$ galaxies\footnote{We
note that the \citet{shen03} early-type sample is {\it not} color selected,
thus, there will be some contamination from centrally-concentrated
galaxies with blue colors.  Yet, Shen et al. have stated that the majority
of $n\ge2.5$ SDSS galaxies have red colors of $(g-r)>0.7$
\citep[see also][]{blanton03d}.}
these authors found a \lr scaling relation in the local universe that is
well fit by a simple power law, and they showed that SDSS
early types of a given luminosity follow a log-normal size distribution
given by
\begin{equation}
f(r_{50},\bar{r}_{50}(L),\sigma(L)) = \frac{1}{\sqrt{2\pi}\sigma(L)} \exp{ \left[ -\frac{\ln^2(r_{50}/\bar{r}_{50}(L))}{2\sigma^2(L)} \right] } \frac{dr_{50}}{r_{50}} ,
\label{lognorm_eq}
\end{equation}
with characteristic median size $\bar{r}_{50}(L)$ and dispersion $\sigma(L)$.
The \citet{shen03} work is based on half-light radii
and total absolute magnitudes from \sersic 
(PSF-corrected) fits to the azimuthally-averaged surface brightness profiles
in the $r$-band \citep[for radial profile fitting see][]{stoughton02}.  
These present-day ($0.05<z<0.15$) sizes are given
in observed $r$-band; nevertheless, at a median redshift of $z\sim0.1$
the sizes are within 1\% of $r$-band rest-frame according to our passband 
correction calculation (\S \ref{sizes}).  Using $k$-corrections from
\citet{blanton03b}, the SDSS galaxy luminosities have been
corrected to the $r$-band rest-frame ($z=0$), which we translate to rest-frame
$V$-band in our adopted ($h$ is free parameter)
cosmology using a mean $(V-r)=0.33$ color for 
E/S0s \citep{fukugita95}.  Therefore, as a function of absolute $V$-band
magnitude $M_V^{\prime}=M_V-5\log_{10}{h}$, the local \lr relation has median
$r$-band size (in units of $h^{-1}$\,kpc)
\begin{equation}
\bar{r}_{50}(M_V^{\prime}) = 10^{(-0.26M_V^{\prime} - 4.93)} ,
\label{medR_eq}
\end{equation}
with dispersion\footnote{This expression fits the dispersion of the \lr
relation for both early and late types to within their respective error bars
\citep[see Fig. 6 of][]{shen03}.} 
that increases with decreasing luminosity as
\begin{equation}
\sigma(M_V^{\prime}) = 0.27 + \frac{0.18}{1+10^{-0.8(M_V^{\prime}+19.78)}} .
\label{dispR_eq}
\end{equation}

Equation (\ref{lognorm_eq}) does not directly allow one to plot the ridgeline
of the \lr relation, because it
does not account for the steeply-declining luminosity function
of early-type galaxies, which at large size will shift the ridgeline
towards fainter luminosities.  To account for this we construct
a large grid of galaxy luminosities and sizes that sample the 
\lr relation of $z\sim0$ early types with a log-normal dispersion in the 
size direction given by Equations (\ref{lognorm_eq}-\ref{dispR_eq}).
We then populate this grid so that the $M_V-5\log_{10}h$ distribution matches 
the local $B$-band luminosity function given in \citet{bell04b}, which we 
transform to rest-frame $V$-band assuming a
typical $(B-V)=0.9$ color for E/S0 types \citep{fukugita95}.
The grid spans $-24<M_V-5\log_{10}h<-16$ in 800 (0.01 mag) cells and 
$0<r_{50}<25$ \hkpc in 500 (0.05 $h^{-1}$\,kpc) cells, for a fine
grid of 400,000 total cells with over 16.9 million \lrd values representative
of the \lr distribution of early-type galaxies in the present-day universe.
We illustrate this mock catalog by plotting its median $M_V-5\log_{10}h$
(with 16-percentile and 84-percentile limits) as a function of $r$-band size
in the left panel of Figure \ref{Mock}.
As can be seen in Figure \ref{Lr8}, the \citet{shen03} ridgeline
provides a reasonable fit to the GEMS \lr relations out to $z\sim0.5$.
At larger redshifts the relations
start to deviate from the local power law, with the largest evolution
appearing for the smallest galaxies.  In any galaxy evolution scenario
we expect that passive luminosity evolution of the stellar populations 
in galaxies must play a role. We use the mock catalog to study the
evolution of the \lr relation with redshift in the next section.

\subsection{Luminosity Evolution at Fixed Galaxy Sizes}
\label{Levol}
We now quantify the evolution seen in Figure \ref{Lr8}, which can be
interpreted in terms of passive evolution of an ancient stellar population.
At a given fixed galaxy size, we measure the
luminosity difference relative to the $z\sim0$ relation as a function of
redshift, accounting for selection effects.  
With the largest sample of size measurements for distant
early types to date, we have the first opportunity to examine the evolution of
these galaxies as a function of their physical size.  To this end we divide
our sample into three broad size bins as shown in Figure \ref{Lr8}, 
each with sample size in excess
of 100, spanning the full range of typical sizes, namely:
({\sc i}) $0.5<r_{50}\le1.0\ h^{-1}$\,kpc; 
({\sc ii}) $1.0<r_{50}\le2.0\ h^{-1}$\,kpc; and 
({\sc iii}) $2.0<r_{50}\le5.0\ h^{-1}$\,kpc.

As we probe to farther distances, the $R_{\rm ap}=24$ mag redshift reliability
cutoff corresponds to ever brighter luminosities (see Figure \ref{Lr8}).  
At this magnitude limit
the completeness of our early-type sample goes to zero.  At brighter
magnitudes the completeness rises rapidly to $>90\%$ in most redshift
intervals. The \combo survey includes a
well-defined completeness map constructed from Monte-Carlo simulations of
the survey and its classification scheme \citep{wolf01b,wolf03,wolf04}. 
Specifically,
for red-sequence galaxies we have completeness factors as a function of
$R_{\rm ap}$ for each redshift interval.  We have demonstrated in \S \ref{sizes}
that the GEMS images of red-selected galaxies are not surface-brightness
limited; therefore, the \combo map provides the selection function for our
early-type sample.  For brevity we denote the magnitude and
redshift-dependent selection function as $S(M_V^{\prime},z)$.  In each
redshift slice of Figure \ref{Lr8} we show schematically the
region of non-zero completeness (open-ended rectangles).  We note that
the $R_{\rm ap}=24$ mag cutoff cannot be exactly depicted in rest-frame $V$-band
owing to the color dependence in completeness and the varying transformation 
between observed $R$-band and rest-frame $V$-band magnitudes.

To account for selection effects using distant cluster samples, previous
studies considered only galaxies brighter than a common absolute magnitude
cut at all redshifts \citep[e.g.][]{pahre96,schade97}.
However, \citet{barger98} argued
that the consistent analysis of a passively-evolving galaxy population must
account for the evolution of the magnitude limit to avoid underestimating
the luminosity evolution.

Our situation is
somewhat different in that we wish to measure luminosity evolution by
direct comparison to the well-defined $z\sim0$ reference points afforded
by the \citet{shen03} study.  Conceptually, we want to passively evolve
the local sample back to higher redshifts and apply the
selection function at each redshift.  In practice, we determine the
corresponding present-day luminosity for each GEMS galaxy of a given size using
a Monte-Carlo sampling of the mock local distribution that has been
subjected to the same (but passively faded) selection function as 
our observations. Therefore, for each GEMS galaxy with
$z$, $r_{50}$, and $M_{V,z}-5\log_{10}h$,
we calculate the luminosity difference
$\Delta M_V = M_{V,z}-\langle M_{V,0}\rangle$, where
$\langle M_{V,0}\rangle -5\log_{10}h$ is the average
absolute magnitude from 100 galaxies of the same size drawn randomly
from the mock $z\sim0$ catalog using a passively-faded version of
$S(M_V^{\prime},z)$.
We note that this selection imitates the GEMS galaxy selection while
accounting simultaneously for the passive increase in luminosity that a
present-day galaxy would have at a given epoch. Moreover,
our method is identical to passively evolving (brightening)
the $z\sim0$ mock dataset to the mean redshift of each epoch, applying
$S(M_V^{\prime},z)$, and computing $\Delta M_V$.

To quantify passive fading, we adopt the luminosity
evolution of a single-burst stellar population, with formation redshift
$z_{\rm form}=3$ and metallicity [Fe/H] $=-0.2$, using the
\pegase model \citep{fioc97}.  Below we discuss other formation redshifts
and metallicities, including the null no-passive-evolution case.
In each redshift interval of 
Figure \ref{Lr8} we show the amount that passive evolution shifts the 
$z\sim0$ selection function relative to $S(M_V^{\prime},z)$ by the 
isolated horizontal line.
Since our selection function is redshift-dependent, it is computationally
easier to apply a passively-faded version of $S(M_V^{\prime},z)$
to the $z\sim0$ mock distribution at each epoch, rather than evolving the
entire mock catalog at each epoch.

\begin{figure*}
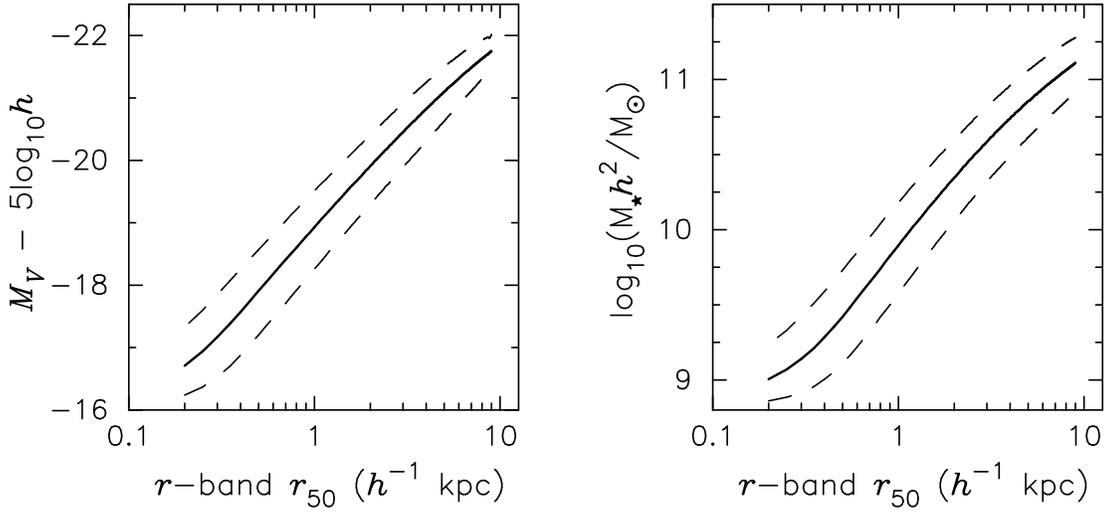

 \centering
 \mbox{
  \includegraphics[scale=0.85, angle=0]{mcintosh_fig7a.ps}
  \hspace{0.75cm}
  \includegraphics[scale=0.85, angle=0]{mcintosh_fig7b.ps}
 }
\caption[]{Mock $z\sim0$ early-type ($n\ge2.5$) \lrd (left) and \mrd
(right) distributions from the
\citet{shen03} log-normal size distributions.  In each panel we plot
the median (solid line) plus 16-percentile and 84-percentile limits
(dashed lines) for
the absolute $V$-band magnitude and stellar mass distributions at
fixed intervals of $r$-band half-light physical size.
The medians in each panel represent the \lr and \mr ridgelines
that we plot in Figures \ref{Lr8} and \ref{Mr8}, respectively.
The \lr relation has luminosity values that are distributed according
to the luminosity function of SDSS-selected, red-sequence 
galaxies from \citet{bell04b}.  Likewise, the \mr relation
has stellar mass values that follow the early-type, $g$-band-derived stellar
mass function from \citet{bell03b}.
\label{Mock}}
\vspace{-0.2cm}
\end{figure*}

\begin{figure*}
\center{\includegraphics[scale=0.85, angle=0]{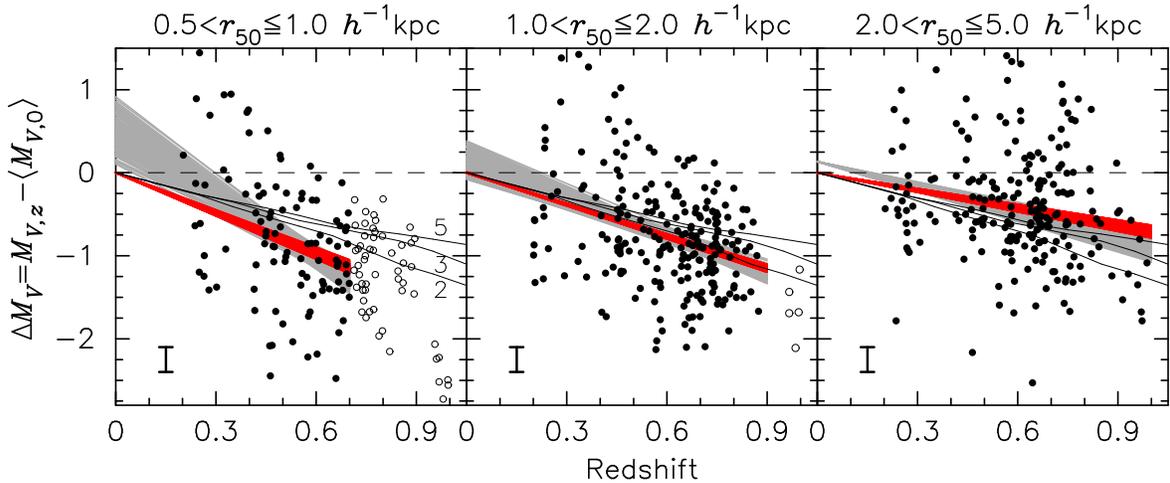}}
\caption[]{Luminosity evolution of early-type galaxies
at fixed size over the last half of cosmic history.  
In each panel, we plot the luminosity difference $\Delta M_V$
between a given GEMS galaxy at redshift $z$ and the average luminosity
($\langle M_{V,0}\rangle -5\log_{10}h$) of a $z=0$ galaxy with the
same size, accounting for selection effects (see text),
against redshift for all galaxies divided into three size bins as follows:
$0.5<r_{50}\le1.0\ h^{-1}$\,kpc ({\sc i});
$1.0<r_{50}\le2.0\ h^{-1}$\,kpc ({\sc ii}); and
$2.0<r_{50}\le5.0\ h^{-1}$\,kpc ({\sc iii}).  We fit a line to
the data, constrained to intercept the origin ($z=0, \Delta M_V=0$), for
galaxies within redshifts limited by the fraction ($>50\%$) of the mock
$z\sim0$ catalog that we include in our luminosity evolution analysis:
$z\le0.7$ ({\sc i}); $z\le0.9$ ({\sc ii}); and $z\le1.0$ ({\sc iii}).  For the
linear regression we depict the included (solid) and the excluded (open)
galaxies separately. The sloped red band represents the best-fit line and
its $1\sigma$ dispersion from bootstrap resampling of the data (see also Table
\ref{LineFits}). We repeat the linear regression of luminosity evolution for
the case where the fit has {\it not} been constrained to go through the origin
(light grey band). The unconstrained fit results are presented in Table
\ref{LineFits_un}. The thin sloped lines are the predicted luminosity evolution
of single-age stellar populations (computed assuming a \citet{kroupa93} IMF
using the \pegase code) with different formation redshifts and metallicities:
$z_{\rm form}=2$, [Fe/H] $=-0.1$; $z_{\rm form}=3$, [Fe/H] $=-0.2$;
and $z_{\rm form}=5$, [Fe/H] $=-0.3$.  The dashed horizontal line
represents no luminosity evolution.  The error bars at lower left of each
panel represent the $\le 15\%$ average uncertainties in COMBO-17 $V$-band
magnitudes.
\label{DelLevol}}
\vspace{-0.2cm}
\end{figure*}

In Figure \ref{DelLevol} we plot the luminosity evolution of the
early-type population in three separate galaxy size bins.
The distributions of $\Delta M_V$ values with redshift show evolution
such that most galaxies of a given size are brighter at larger look-back times.
The scatter in the data is similar to the dispersion given by the log-normal 
distribution from \citet{shen03}, as shown in the left panel of Figure 
\ref{Mock}.
We fit a line (using ordinary least-squares linear regression) to
the $\Delta M_V$ evolution in each panel of Figure \ref{DelLevol},
constrained to intercept the origin ($z=0, \Delta M_V=0$).
We limit this analysis to GEMS galaxies (solid circles) within each size
bins out to a maximum redshift where the selection function is 
$\ge50\%$ according to the mock $z\sim0$ catalog for a passively fading
scenario.  Therefore, as outlined with open rectangles in Figure \ref{Lr8},
we do linear fits to galaxy samples from the three size bins as follows:
({\sc i}) 102 with $z\le0.7$; ({\sc ii}) 227 with $z\le0.9$; and
({\sc iii}) 222 with $z\le1.0$.  We exclude 57 small ({\sc i}) and
7 medium ({\sc ii}) size early types (open circles in Figure \ref{DelLevol}).
In each panel,
the sloped red band gives the linear fit to the luminosity evolution with
$1\sigma$ dispersion.  The regression dispersion represents
the 68-percentile distribution of linear fits to 200 bootstrap resamples 
of the $(\Delta M_V,z)$ values.  We find significant luminosity evolution
for early-type galaxies of all sizes, from $\sim1.0$ mag for small
($0.5<r_{50}\le1.0\ h^{-1}$\,kpc) systems
between $z=0.7$ and now, to $\sim0.7$ mag for the largest galaxies
between $z=1.0$ and the present time.  We note that the apparently large change
in \lr slope with redshift found in Figure \ref{Lr8} is due mostly to our
selection function cutting lower luminosity galaxies at higher redshift.
Nevertheless, once we correct for selection effects, the differing
amounts of luminosity evolution in each size bin of
Figure \ref{DelLevol} does represent a significant
change in \lr slope with redshift not observed in previous studies.

In Table \ref{LineFits} we tabulate the slopes of the constrained linear
fits to the luminosity evolution for each size bin.
We repeat this analysis using passive fading of $S(M_V^{\prime},z)$ that
is slightly stronger ($z_{\rm form}=2$, [Fe/H] $=-0.1$) and
slightly weaker ($z_{\rm form}=5$, [Fe/H] $=-0.3$)\footnote{
In both cases the metallicities are changed in concert with the 
formation redshifts to reproduce the observed color
evolution of the red sequence; see e.g., \citet{bell04b} for 
more details}.  We see in
Table \ref{LineFits} that the measured luminosity evolution is statistically
equivalent ($<1\sigma$ differences) between the analyses using three
different passive fading parameters.  In addition,
we calculate $\Delta M_V$ values under the null hypothesis of no passive
evolution.  As \citet{barger98} pointed out, not including passive evolution
of the luminosity cutoff (here defined by our completeness map) will
underestimate the evolution.  Nonetheless, even under the null condition
we still find significant luminosity evolution in early-type galaxies of
a given size.  Furthermore, we find that the amount of evolution continues 
to vary between different fixed-size galaxy populations.

Last, we compare the constrained fit to the luminosity evolution with 
a fit that is {\it not} constrained to go through the $z=0$, no evolution 
origin.  For this test we use the same $\Delta M_V$ data based on the
mock catalog selection with $z_{\rm form}=3$ passive fading assumed, and
the same method including bootstrap estimation of dispersion.
We plot the unconstrained fits with light grey bands in each panel of
Figure \ref{DelLevol},
and we provide the best-fit linear parameters (slope and $z=0$ intercept)
in Table \ref{LineFits_un}.  We point out that the intrinsic scatter
in the $\Delta M_V,z$ correlation, and the fairly restricted redshift
range over which we fit, both combine to produce a less well-constrained
fit for the smallest galaxies compared with the two larger size bins.
In all cases, significant fading of the early-type
galaxy population at a given size is clearly detected, and the difference in
fading rates between small and large galaxies is preserved independent of
assumptions regarding passive evolution and fitting method.

The two methods of fitting luminosity evolution have different merits.
The constrained fitting is anchored directly to $z=0$, which allows the
use of our full knowledge of the local universe.
Furthermore, the constrained fits are quite robust as illustrated by
the very narrow dispersion we find with bootstrap resampling.
On the other hand, the unconstrained fits show clearly the differences
between GEMS and SDSS photometric and size measurement calibrations,
and for slight deficiencies in our treatment of selection effects;
the $z = 0$ offset is $\la 0.25$ mag in all bins.

\subsection{Stellar Mass--Size Relation}
\label{MrRel}
\begin{figure*}
\center{\includegraphics[scale=0.75, angle=0]{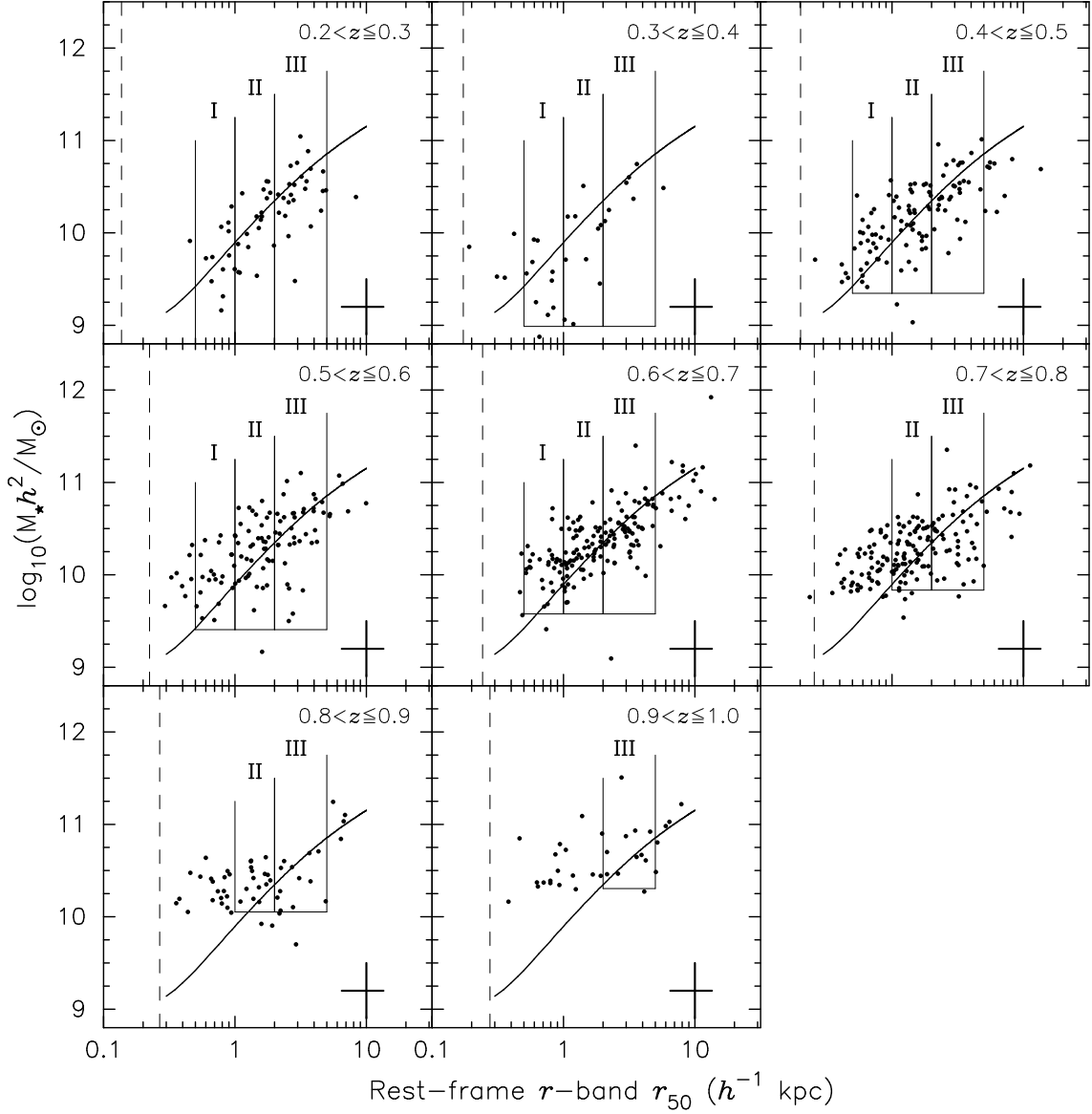}}
\caption[]{{\it Stellar mass--size} relations for 728 early-type galaxies 
with $0.2<z\le1.0$ split into eight redshift slices as in Figure \ref{Lr8}.
In each panel we plot the physical size $r_{50}$ ($z$-band corrected to
$r$-band)
against the COMBO-17 SED-based stellar
mass.  The error bars shown at the bottom right of each panel give the
35\% $r_{50}$ and the factor of two (0.3 dex) 
random M$_{\star}$ uncertainties.
Using our mock M$_{\star},r_{50}$ catalog 
(see right panel of Figure \ref{Mock}), we show the
\mr relation ridgeline for $z\sim0$ early types ($n\ge2.5$) from
\citet{shen03} in each redshift slice for comparison (diagonal line).
Following the format of Figure \ref{Lr8},
we delineate the three fixed-size bins ({\sc i}, {\sc iii}, {\sc iii})
with vertical lines, and we depict the low-mass cutoff (horizontal lines)
corresponding to the limit for reliable \combo redshifts ($R_{\rm ap}=24$ mag).
Thus, the open-ended rectangles outline the
non-zero completeness regions of M$_{\star},r_{50},z$-space.
The physical size corresponding to the PSF radius ($0\farcs05$)
at each redshift is shown by the vertical dashed line.
\label{Mr8}}
\vspace{-0.2cm}
\end{figure*}

We turn our attention now to exploring the stellar mass evolution of GEMS
early-type galaxies as a function of fixed size.
In Figure \ref{Mr8} we present the \mr relations for 728
early types in eight $\Delta z=0.1$ redshift bins spanning $0.2<z\le1.0$.
As with \lrns, the \mr relations are quite apparent to at least $z\sim0.8$.

In each panel of Figure \ref{Mr8} we display the \mr relation
ridgeline for $z\sim0$ early-type galaxies from SDSS.
\citet{shen03} showed that these
galaxies also follow a log-normal size distribution of the same general
form as Equation (\ref{lognorm_eq}) with a stellar mass dependence.
These authors derived stellar masses from SDSS Petrosian luminosities
using a model-dependent stellar \mlstar from 
\citet{kauffmann03}\footnote{\citet{bell03b} showed that spectrally-derived
and color-derived stellar masses are the same to within 30\%.}.
For our adopted ($h$ is free parameter) cosmology, the local \mr relation has
characteristic median size 
\begin{equation}
\log_{10}{(\bar{r}_{50}/h^{-1} {\rm kpc})} = 0.56\log_{10}{({\rm M}_{\star}h^2/{\rm M}_{\sun})} - 5.52
\label{medR_M_eq}
\end{equation}
with log-normal dispersion
\begin{equation}
\sigma({\rm M}_{\star}h^2) = 0.29 + \frac{0.53\times10^{10}{\rm M}_{\sun}}{{\rm M}_{\star}h^2} ,
\label{dispR_M_eq}
\end{equation}
given in observed $r$-band size (S. Shen 2004, private communication).
Recall that we show in \S \ref{LrRel} that these sizes are within 1\% of
rest-frame $r$-band.

Following the method in \S \ref{LrRel},
we create a mock catalog of \mrd values that represent
the \mr distribution of early-type galaxies at $z\sim0$ from
\citet{shen03}.  Briefly, we construct a 400,000 cell grid of galaxy
stellar masses and sizes that follow the log-normal distribution
given by Equations (\ref{medR_M_eq} \& \ref{dispR_M_eq}).
The grid spans $8.8<\log_{10}({\rm M}_{\star}h^2/{\rm M}_{\sun})<12.8$ 
in 800 (0.005 dex) cells, and $0<r_{50}<25$ \hkpc in 500 (0.05 $h^{-1}$\,kpc) 
cells.  We populate this grid to match the $g$-band-derived
stellar mass function
for local early-type galaxies from \citet{bell03b}.  This mock catalog
contains over 17.4 million M$_{\star},r_{50}$ values.
We use this mock catalog to study the stellar mass evolution of
early types as a function of redshift in the next section.
In the right panel of Figure \ref{Mock}, we plot the median value of
$\log_{10}{({\rm M}_{\star}h^2/{\rm M}_{\sun})}$, with its 16-percentile and
84-percentile limits, 
against $r$-band size.  This median provides the \citet{shen03}
\mr ridgeline that we compare to the observed relations
at each redshift interval in Figure \ref{Mr8}.  Here we see
that sizes and stellar masses of the GEMS
early-type galaxies follow a correlation that is generally consistent
with the local \mr relation out to $z\sim0.8$, where
the \combo redshift reliability limit begins to cut into the observed
relation.  Under the assumption of simple passive luminosity evolution,
we expect that galaxies of a given size will maintain 
a constant stellar mass.  We investigate this expectation in the next section.

\subsection{Stellar Mass Evolution at Fixed Galaxy Sizes}
\label{Mevol}
To directly compare any evolution in
the luminosity and stellar mass of early types we must analyze the GEMS
\mr relation in an analogous manner to the \lr relation.
As with our luminosity evolution analysis of distant galaxies, we
calculate how much the \mr relation shifts with respect
to the local \citet{shen03} relation, while accounting simultaneously for
the selection effects of our observations.  Just as the \combo redshift
reliability limit at $R_{\rm ap}=24$ imposes an effective absolute magnitude cut
at each epoch, the minimum stellar mass that a galaxy has in our sample 
increases with redshift as expected in a magnitude-limited sample
(horizontal lines in each panel of Figure \ref{Mr8}).
Therefore, for each redshift bin we determine the simple linear relation
between M$_{\star}$ and $R_{\rm ap}$ for each galaxy to convert the
\combo completeness map into a redshift-dependent selection function given
as a function of logarithmic stellar mass; i.e. 
$S(\log_{10}{\rm M}_{\star}^{\prime},z)$, 
where ${\rm M}_{\star}^{\prime}=({\rm M}_{\star}h^2/{\rm M}_{\sun})$.
We apply $S(\log_{10}{\rm M}_{\star}^{\prime},z)$ to the $z\sim0$ mock catalog
as before, with one important difference -- there is no need to correct the
selection function for the effects of passive evolution because a
passively-evolving galaxy population is {\it not} expected to evolve in 
stellar mass.

We divide the sample into the same three fixed-size bins and for each GEMS
early-type galaxy we calculate the stellar mass difference
$\Delta \log_{10}{\rm M}_{\star}^{\prime} = \log_{10}{\rm M}_{\star,z}^{\prime} - \langle \log_{10}{\rm M}_{\star,0}^{\prime} \rangle$,
which is the difference (in log-space) between the stellar mass of the
galaxy observed at redshift $z$, and the average stellar mass of 100 
galaxies of equivalent
size drawn at random from the $z\sim0$ mock \mrd distribution,
which is weighted by $S(\log_{10}{\rm M}_{\star}^{\prime},z)$.  Therefore,
for each GEMS galaxy of a fixed size we find the present-day stellar
mass from the mock catalog using the same selection as our observations.
The mock catalog is based on a $g$-band mass function; below we discuss
the effects that two other mass function
estimates have on the local sample selection we use for this calculation.

\begin{figure*}
\center{\includegraphics[scale=0.85, angle=0]{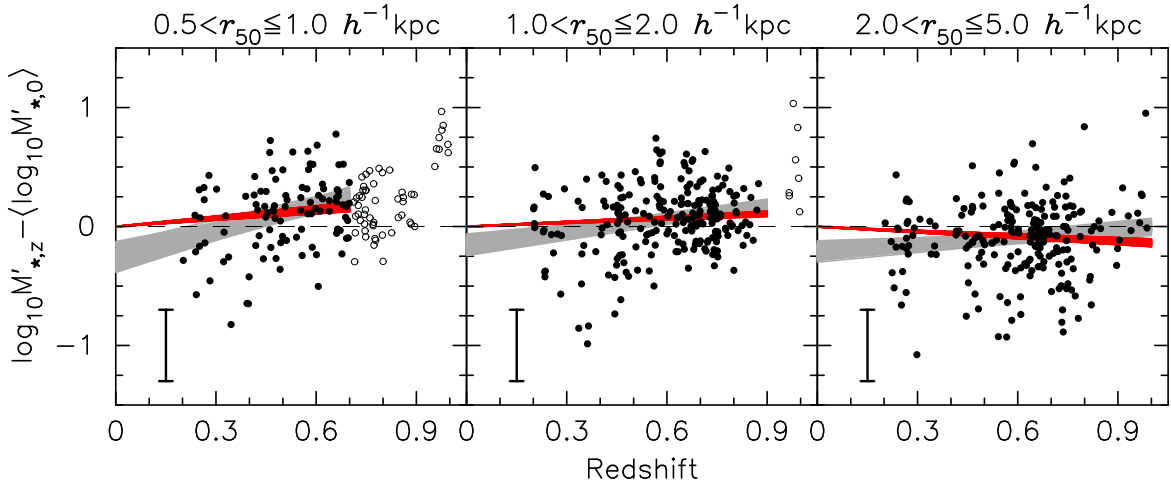}}
\caption[]{Stellar mass evolution of early-type galaxies
with fixed physical sizes over last half of cosmic history.  
In each panel we plot the stellar mass difference
$\Delta \log_{10}{\rm M}_{\star}^{\prime} = \log_{10}{\rm M}_{\star,z}^{\prime} - \langle \log_{10}{\rm M}_{\star,0}^{\prime} \rangle$
(where ${\rm M}_{\star}^{\prime}=({\rm M}_{\star}h^2/{\rm M}_{\sun})$)
against redshift for all galaxies divided into three size bins shown.
$\Delta \log_{10}{\rm M}_{\star}^{\prime}$ represents a measure of
the stellar mass evolution from the $z\sim0$ \citet{shen03} \mr
relation (see text for details).
As with our luminosity evolution analysis (see Figure \ref{DelLevol}),
we perform a linear regression for the stellar mass evolution, both
constrained (red bands) and not constrained (light grey bands)
to intercept
the origin ($z=0, \Delta \log_{10}{\rm M}_{\star}^{\prime}=0$).  For these
fits we use only those galaxies (solid circles) within the following
redshifts: $z\le0.7$ ({\sc i}); $z\le0.9$ ({\sc ii}); and $z\le1.0$ ({\sc iii}).
Furthermore, the set of red and grey bands represent the 68-percentile
distribution of best-fit lines to 200 bootstrap resamplings of the data.
We tabulate the fit results in Tables \ref{LineFits} and \ref{LineFits_un}.
The dashed horizontal line represents no stellar mass evolution.
The error bars at lower left of each
panel represent the 0.3 dex random uncertainties in our M$_{\star}$ estimates.
\label{DelMevol}}
\vspace{-0.2cm}
\end{figure*}

To quantify any redshift evolution of the stellar mass of early-type
galaxies, we fit lines to the $\Delta \log_{10}{\rm M}_{\star}^{\prime}$
values as a
function of redshift following the same procedure as in \S \ref{Levol}.
As before, we limit the analysis to the same subset of GEMS galaxies per
fixed-size bin (see open rectangles marked {\sc i}, {\sc ii}, and {\sc iii}
in Figure \ref{Mr8}).  In the three panels of Figure \ref{DelMevol} we
plot the linear fits to the stellar mass evolution constrained to have
no evolution at $z=0$.  The red bands represent the best-fit and the
68-percentile distribution of fits ($1\sigma$ dispersion) from 200 bootstrap
resamples.  We present the fit results in Table \ref{LineFits}.
We find a modest change in stellar mass for early-type galaxies in each
size bin.  For the two smaller size bins ($\le2 h^{-1}$\,kpc) the stellar
mass appears to have been somewhat larger in the past.  Conversely,
the largest early-type galaxies ($>2 h^{-1}$\,kpc) may have had slightly
less stellar mass at earlier look-back times.  We repeat the above analysis
using local mock \mrd distributions that follow two
additional stellar mass functions based
on $K$-band ($r$-band concentration-selected) and $g$-band
($g-r$ color-selected; i.e., $g_{\rm col}$) from \citet{bell03b}.
As shown in Table \ref{LineFits}, we find that
the stellar mass evolution results are independent of the mass function
we use to construct the present-day mock catalog.

Last, we repeat the linear regression analysis for the redshift evolution of 
$\Delta \log_{10}{\rm M}_{\star}^{\prime}$ ($g$-band-based), but in this
case we allow the fit to be unconstrained.  We present these results in
the panels of Figure \ref{DelMevol} using light grey bands (see also
Table \ref{LineFits_un}).
We see that the unconstrained fits have slopes that are within $\sim2\sigma$ 
of flat; i.e. consistent with no stellar mass evolution for galaxies
of given fixed sizes as expected if they contain passively-evolving stellar
populations as suggested by the luminosity evolution (\S \ref{Levol}).
Furthermore, the intercepts for $\le2$\hkpc galaxies are statistically
equivalent (i.e., within $3\sigma$) to no evolution at $z=0$, thus, this shows
that our stellar mass estimates are remarkably well-calibrated with the local
SDSS stellar masses estimated in an independent manner.  We note that
larger galaxies are about
0.2 dex more massive at $z=0$ than the expectations from SDSS, still
a fairly small offset given the uncertainties of our estimates.

\section{Discussion}
\label{Disc}

\subsection{Comparison with Previous Work}
Overall the evolution that we see for the \lr relation of different galaxy
subsets is in good agreement with earlier, more-restricted studies based on 
much smaller data sets. The large field of the GEMS survey 
($\sim 15\,h_{100}^{-1}$ Mpc on a side) permits us to come closer to a
cosmologically-representative ``field'' sample of early-type galaxies.
Additionally,
the large sample size allows us for the first time to demonstrate that
the strength of luminosity evolution, at a given size, depends inversely
on early-type galaxy size:
$\Delta M_V / \Delta z \sim -1.6\pm0.1$ for small galaxies
($r_{50} \le 1 h^{-1}$\,kpc),
$\Delta M_V / \Delta z \sim -1.3\pm0.1$ for medium-sized galaxies
($1<r_{50} \le 2 h^{-1}$\,kpc), and
$\Delta M_V / \Delta z \sim -0.7\pm0.1$ for large galaxies
($2<r_{50} \le 5 h^{-1}$\,kpc).

Past work has been limited mostly to cluster environments
\citep{barrientos96,pahre96,schade97,barger98,ziegler99},
where the increased densities provided reasonable sample sizes ($N<200$)
within the small imaging field of WFPC2.  These studies all
found an increase in luminosity with look-back time (at a given size)
of about 1 magnitude in rest-frame $B$-band per unit redshift. Allowing
for the $B$ versus $V$-band differences and comparing
the evolution at comparable sizes ($\sim 3 h^{-1}$\,kpc), we find
in comparison no evidence for environmental differences in evolution.

\citet{schade99} and \citet{trujillo04c} have selected small
early-type galaxy samples ($N<50$) with WFPC2 observations that are
sampled from ``field'' environments at $z<1$ in
a way that should be statistically representative of the cosmic average,
and found similar luminosity evolution for galaxies of a fixed size in
the field as compared to clusters.
For example, Trujillo \& Aguerri found that small early-type galaxies in the
HDFs were $1.35\pm0.1$ mag brighter in rest-frame $V$-band at $z\sim0.7$
when compared to the present-day relation from \citet{shen03}.

Taking into account the changes in stellar \ml that are implied by
the color evolution of early-type galaxies in the last 8 Gyr, we
find that the largely passive luminosity evolution of the \lr relation
implies little or no evolution of the \mr relation since
$z \sim 1$. Figure \ref{DelMevol} 
shows that the average stellar mass in early-type
galaxies of a given size has remained fairly constant over the
interval $0<z<1$. Concentrating on the constrained fits (see
also Table \ref{LineFits}), we place an upper limit on the change in the \mr
relation of 0.25 dex, in the sense that galaxies may have been slightly more
massive at a given size at $z\sim 1$ than they are at the present.
Our findings agree with those of \citet{trujillo04a},
who found very little stellar mass evolution since $z\sim2.5$ using a
sample of 168 galaxies of all morphological types from the HDF-South. In
addition, we find that early-type galaxies of a given fixed size had
similar stellar masses at $z\sim1$ as they do today. This is qualitatively
consistent with studies of the Kormendy relation
\citep[e.g.,][]{barger98}, absorption lines
\citep[e.g.,][]{ziegler97,kelson01}, color--magnitude relation
\citep[e.g.,][]{kodama99,bell04b,holden04}, and FP
\citep[e.g.,][]{kelson00c,vandokkum01a,treu02,
vandeven03,gebhardt03,vanderwel04}, who all found changes in \ml consistent
with passive evolution of the stellar populations in early-type galaxies
since $z\sim1$.

The unconstrained fits to the data in Figure \ref{DelMevol} 
may indicate marginally-significant
evolution.  In all cases, the observed trends within the redshift
ranges probed by the data are 0.5 dex or less, in the sense
that distant early-type galaxies may be denser than their low-redshift
counterparts.  This trend appears strongest for the 
$0.5 < r_{50} \le 1.0 h^{-1}$\,kpc sample, where we find stronger evolution
of the \lr relation (Figure \ref{DelLevol}, left panel). At this stage, it
is difficult to reliably assess the significance of this possible 
evolution owing to systematic uncertainties of the stellar 
\ml ratios (which are $\sim 0.3$ dex).  In this context, it is interesting
to note first results of a systematic study of the 
$0.7<z<1.1$ early-type galaxy FP
by \citet{vanderwel04}; they found that dynamically-derived
\ml ratios of six intermediate redshift early-type galaxies
showed a slightly stronger color--\ml correlation than expected
in a single burst model.  The most straightforward interpretation of 
this would be an increasing importance of bursts of 
star formation in low mass (therefore small) early-type
galaxies; aging bursts bias color-based \ml estimates
high by up to 0.3 dex \citep{bell01}.  Thus, the 
apparently strong evolution of the \mr relation for small
(low mass) galaxies could be enhanced artificially by 
the effect of bursts of star formation, whose frequency
is naturally expected to be higher at earlier times.   

\subsection{Understanding the Evolution of Early-Type Galaxy
Scaling Relations}
\label{disc2}

Taken by themselves, the observed evolution of the \lr and \mr
scaling relations is insufficient to constrain the characteristic
evolutionary fates
of {\it individual} early-type galaxies. For example, one can imagine
morphological transformations and changes in star formation history caused
by galaxy mergers, disk re-growth, and fading of previously star-forming
disks that will cause galaxies to drift in and out of the early-type
and/or red-sequence galaxy populations \citep[e.g.,][]{baugh96,steinmetz02}.
Thus, evolution of the scaling relation of the populations
should be interpreted in terms of the evolution of the early-type galaxy
populations, rather than in terms of the evolution of the individual
galaxies themselves.

Observationally, the results of this paper are consistent with passive
evolution of the early-type galaxy population as predicted by the
monolithic collapse scenario \citep{eggen62,larson74}. This model
describes the formation of present-day spherical systems through the
collapse of a massive gas cloud followed by a brief burst of SF early in
the history of the universe ($z_{\rm form}>2$). For this model any changes
in the observed properties of early-type galaxies over time are due to
simple passive fading of the coeval stellar populations.

Yet, a large number of works \citep{chen03,bell04b,drory04,conselice04,cross04}
have found that the total stellar mass density in early-type galaxies,
defined either by color or morphology, has built up by roughly a factor of
two in the last 8 Gyr since $z \sim 1$. Regardless of the mechanisms
driving this evolution (e.g., mergers, disk fading, etc), the lack
of drastic evolution in the \mr relation indicates that early-type
galaxies to first order 
either {\it move along} the \mr relation as they evolve, or they
{\it appear on it} when they join the sample of early-type red sequence 
galaxies. The lack
of strong evolution in the \mr relation since $z \sim 1$
is an important constraint that early-type galaxy formation theories will
have to satisfy.

It is interesting to note that both disk fading and galaxy mergers may
naturally satisfy the observational constraints. 
It is possible that some disk-dominated
galaxies with reasonably massive bulges cease to form stars at
intermediate redshift (owing perhaps to gas consumption or removal of
its gas supply). As
the massive stars in the disk die, the disk fades very quickly, increasing
greatly the prominence of the bulge. Under the assumption that the disk
{\it stellar mass--size} correlation does not evolve with redshift 
\citep[as justified by][Barden et al., in preparation]{trujillo04a}, and noting
that the local disk and bulge {\it stellar mass--size} correlations are within
0.2 dex of each other over a wide stellar mass range \citep[see Fig.\ 11
of][]{shen03}, it is quite possible that disk fading would produce
early-type galaxies that adhere closely to the redshift-independent
\mr relation.

Another possible formation mode for early-type galaxies is through major
galaxy mergers \citep[e.g.,][]{naab03,khochfar03}, where
the remnant has suffered violent relaxation and is
spheroidal and pressure-supported \citep{toomre72,barnes92}.
Detailed studies of close pairs
have shown that an important fraction of $\sim L^*$ galaxies may merge
between $z \sim 1$ and the present day
\citep{carlberg94,lefevre00,patton02,conselice03}, making this a
potentially important formation mode for early-type galaxies. Yet, it is unclear
whether a merger between two gas-rich galaxies will lead to a remnant
that will satisfy our observed lack of significant evolution in the
\mr correlation \citep[see e.g.,][for a systematic study of gas-rich galaxy
mergers; unfortunately the relationship between progenitor and remnant size
was not explored in this work]{barnes02}. 

Mergers between gas-poor progenitors are easier to
model, and numerous studies have found that merging early-type galaxies
will produce remnants that adhere reasonably well to the 
FP, but slowly drift away from the \mr relation of the progenitor population
\citep{navarro90,dantas03,nipoti03,gonzalez03,shen03}.  \citet{dantas03},
\citet{nipoti03} and \citet{navarro90} showed that
$\log_{10}(r_{\rm remnant}/r_{\rm original})\sim 1.2 
\log_{10}({\rm M_{remnant}/M_{original}})$; that is, one
dissipationless 1:1 merger will lead to a size increase of 
$\sim 0.35$ dex.  For a factor of two increase in stellar
mass, the \mr relation (e.g., Equation \ref{medR_M_eq}) shows 
a $\sim 0.15$ dex increase in size.
Thus, galaxies undergoing dissipationless merging
will gradually move towards radii that are larger than galaxies
not undergoing merging.  As outlined earlier,
dissipationless merging is expected to be a much more 
important process for high-mass galaxies \citep{khochfar03}.
Therefore, we can focus on the evolution of the \mr relation 
for galaxies with $2 < r_{50} \le 5 h^{-1}$\,kpc --- 
here the observations place an upper limit
of 0.25 dex on the evolution of the \mr relation zero point
since $z \sim 1$, and show a scatter of $\sim 0.5$ dex.  
Therefore, combining the model predictions (that one major merger
will move a remnant 0.2 dex from the \mr relation defined
by the progenitor population) and the observational
upper limits on \mr relation evolution and scatter, we conclude
that the most massive galaxies have suffered at most one
major dissipationless merger since $z \sim 1$ on average (from 
the zero point evolution), 
and that only a small fraction of massive early-type galaxies could
have suffered several major dissipationless mergers (from the scatter).

\subsection{Strength of Luminosity Evolution Dependence on Galaxy Size}
\label{disc3}

It is worth discussing briefly the size dependence in the evolution of the
\lr relation apparent in Figure \ref{DelLevol}. Focusing on the largest
galaxies with $2<r_{50}\le5\ h^{-1}$\,kpc, we find that the intercept of
the \lr relation is $\sim 0.8$\,mag brighter at $z \sim 1$,
compared to the present day. In contrast, smaller galaxies evolve more
rapidly towards the present; extrapolated to $z \sim 1$, we find 1.5--2
mag of evolution in the \lr relation.

This difference in evolution is statistically significant, thus leaving
two possible classes of interpretation: {\it (i)} smaller galaxies have
significantly younger luminosity-weighted stellar ages, leading to
significantly more rapid luminosity evolution; and {\it (ii)}
there is scale-dependence in the evolution of the \mr
relation related to different formation routes for low- and high-mass
early-type galaxies. At some level, both effects should contribute; we
will conclude here that we cannot at this stage differentiate between
these two possibilities.

The interpretation that smaller galaxies have younger stars compared to
the larger systems, is consistent with
recent deep determinations of galaxy cluster CMRs at
$0.8 \la z \la 1$ \citep{kodama04,delucia04}, where the
faint end of the red sequence is systematically suppressed compared to the
local universe\footnote{It is worth noting that one may well expect a
change in CMR slope with redshift if low-mass
galaxies are on average younger.}. This is also consistent with analyses
of the stellar populations and dynamically-derived \ml ratios of
morphologically-selected early-type galaxies in the local universe
\citep{kuntschner00,trager00,thomas03}, and at $z\sim 1$
\citep{vanderwel04}, where low-mass early-type galaxies have
younger stellar populations than high-mass early types. In this
interpretation, one expects that the bulk of the slope evolution in the \lr
relation is driven by the rapid evolution in \mlstar
of low-mass galaxies, compared to their older high-mass
counterparts\footnote{It is interesting to speculate that owing to biased
galaxy formation (in the sense that the central parts of high-mass halos
collapse at higher redshift than the central parts of lower-mass halos),
one may expect a mild dependence of age on mass, under the assumption that
one can suppress late cooling in early-type galaxy halos. One may expect
to see an age difference between cluster and field early-types in this
interpretation; the observational evidence in this regard is
controversial, and the most fair statement that can be made is that if there
is an age difference as a function of environment, then this dependence is
rather weak \citep[e.g.,][]{bernardi98,hogg03}. }.

There is also evidence from studies of local early-type galaxies that
low-mass early-types have different properties than high-mass early-type
galaxies \citep{kormendy96,gebhardt96,faber97,ravindranath01}.
Stereotypical high-mass early-type galaxies have boxy isophotes,
steep outer light profiles, constant surface brightness
(or ``cuspy'') cores, and are supported
primarily by random motions of stars in a triaxial potential. In contrast,
lower-mass early-type galaxies tend towards diskier isophotes,
steep power-law light profiles without a resolved core,
and derive partial support from organized rotation. 
These differences in properties are
naturally interpreted in terms of different formation mechanisms, where
higher-mass early-type galaxies suffer from late dissipationless major
mergers and lower-mass early-type galaxies result from lower mass-ratio
interactions and/or mergers of galaxies with significant gas contents
\citep{naab03,khochfar03}.  It is also possible that 
some low-mass early-type galaxies could form through 
disk instabilities in small, very high surface density disks
\citep[see e.g.,][for one implementation of this process]{cole00}.
Interpreted in this way, the evolution of the slope of the \lr
relation could be attributed primarily to genuine
evolution in the \mr relation, and differences in the rate
of \mlstar evolution as a function of stellar mass may play a
secondary role.

Viewed in this context, the slight differences in stellar mass evolution
(Figure \ref{DelMevol}) as a function of size suggests 
that it is impossible to rule
out scenario {\it (ii)} at this stage. Yet, the earlier discussion made it
clear that within the uncertainties inherent to the estimation of \mlstar
from optical SEDs, it is impossible to convincingly argue against or
in favor of changes in the \mr relation. Therefore, we
conclude that disentangling the relative importance of the above two
mechanisms in driving the redshift-dependent slope of the \lr
relation is impossible at this time. Future works, utilizing larger
samples with spectroscopic redshifts and velocity dispersions, will
clarify this issue considerably.

\section{Conclusions}
In order to place constraints on the processes driving
early-type galaxy evolution at $z\le1$, we have analyzed the evolution
of the {\it luminosity--size} and {\it stellar mass--size} relations using a
sample of 728 early-type galaxies with $0.2<z\le1.0$ from the GEMS
survey in the extended CDFS.
The sample was selected to have concentrated
light profiles with \sersic $n\ge 2.5$ and rest-frame optical colors
within 0.25 mag of the red-sequence ridge at its epoch.
At a given half-light radius $r_{50}$, early-type galaxies were more
luminous in the past. Selection effects were carefully accounted
for throughout using detailed completeness maps derived
for the parent COMBO-17 photometric redshift catalog.
We find that the \lr relation has evolved
in a manner that is consistent with the passive aging of ancient stellar
populations. In addition, we find evidence for an evolving tilt of the \lr
relation, in the sense that smaller galaxies fade more rapidly
towards the present day than larger galaxies. Using stellar
mass-to-light ratios derived from the COMBO-17 SEDs, we rule out
any substantive evolution in the \mr relation, also consistent with
the simple passive evolution scenario.

Clearly, these results would be consistent with passive evolution
of the early-type galaxy population (i.e., passive aging of existing stars, 
no merging), where a younger age for
smaller galaxies is indicated by the tilting \lr relation.
Yet, this interpretation is too simplistic; a number of surveys have
demonstrated that the total stellar mass in the early-type galaxy
population (as defined here) 
has increased by roughly a factor of two since $z \sim 1$.
Bearing in mind this evolution, our results imply that newly-added
early-type galaxies follow \lr and \mr relations similar to
the more established early types.  A disk-fading origin for 
the newly-added early-type galaxies appears to be consistent with 
the data.  Through comparison with models, the 
role of dissipationless merging is limited to $<1$
major merger on average since $z = 1$ for the most massive galaxies.
The predicted evolution for gas-rich mergers is
not yet robustly predicted, so it is impossible to comment
on this possible formation route.  In this context, the evolving
tilt of the \lr correlation could reflect a different origin
of low-mass early-type galaxies and/or younger stellar populations;
the present data is insufficient to discriminate between these possibilities.

\acknowledgements  
We are extremely grateful to Shiyin Shen for providing $r$-band 
log-normal size distributions as a function of stellar mass, and
for helpful discussions regarding his all-important local scaling
relations from SDSS.
Support for the GEMS project was provided by NASA through grant number GO-9500
from the Space Telescope Science Institute, which is operated by the
Association of Universities for Research in Astronomy, Inc. for NASA, under
contract NAS5-26555.
DHM and SJ acknowledge support from the National Aeronautics
and Space Administration (NASA) under LTSA Grant NAG5-13102 (DHM)
and NAG5-13063 (SJ)
issued through the Office of Space Science.
EFB and SFS acknowledge financial support provided through
the European Community's Human Potential Program under contract
HPRN-CT-2002-00316, SISCO (EFB) and HPRN-CT-2002-00305, Euro3D RTN (SFS).
CW is supported by a PPARC Advanced Fellowship. CH acknowledges support 
from GIF.
Thanks for useful discussions - Rose Finn, Kelly Holley-Bockelmann, 
Neal Katz, Dusan Keres, Ari Maller, Houjon Mo, Ignacio Trujillo.

\appendix
\section{\sersic Fits}
\label{fitting}

For our primary analysis we use \gim 
\citep[Galaxy IMage 2D,][]{simard99,simard02} to fit
a single \sersic model to the two-dimensional radial light profile of 
each galaxy from the F850LP imaging. The \gim software takes
into account the point-spread function (PSF) by convolving it with
the best-fit model profile during fitting, and \gim uses the SExtractor-produced
segmentation mask to deblend galaxies from nearby companions.
The profile fits provide
information that allow us to select morphologically early-type galaxies
from the red sequence (as described in \S \ref{QMorph}).  In addition,
effective size (i.e. half-light radii $r_{50}$; see \S \ref{sizes}) 
measurements and other
structural parameters are provided by the \sersic fitting.

Galaxy profile fitting requires a well-defined model of the PSF and
precise knowledge of the background sky level.
We use a universal, high S/N PSF derived from 548 bright but
unsaturated stars from the Mark I suite of GEMS ACS frames in
the F850LP passband 
\citep[for a first impression see][plus Jahnke et al., in preparation]{jahnke04}. Through
detailed testing we find that this PSF provides sufficient accuracy for the 
galaxy fitting (see H05).  Moreover, with our simulations we
explore three different methods for determining sky values:
(1) using the SExtractor sky value ``local'' to each galaxy;
(2) adopting a fixed constant value per individual ACS frame given by the
SExtractor ``global'' estimate; and (3) letting \gim calculate the mean sky.
We find that the \sersic fitting results are most robust when using 
the first method, bearing in mind that other local sky estimates may
perform equally well or better.

The \sersic model is represented by seven free parameters: total
intensity, semi-major axis scale length, ellipticity $e$,
position angle, index $n$, and model $x,y$ centering.  For each galaxy
fit, \gim automatically determines the initial values and limits for 
the parameter space to be explored by using the segmentation mask,
which is affected by the choice of detection
configuration (see \S~\ref{hstimaging}).  With our simulations we find
that this method provides reliable fits for galaxies with surface
brightnesses of $\mu_{50}<24$ mag asec$^{-2}$.
As can be seen in Figure \ref{Obsvd}, all red-sequence galaxies are
brighter than this limit.

We check the fits using a variety of tests.
First, we check the recovered ellipticity of each fit and find no trends
with redshift or total $m_{850}$ magnitude.  Likewise, the \sersic
$n$ distributions show no trends with $z$ or $m_{850}$ for galaxies
with $n\ge 2.5$.  As expected, we see a slight trend such that highly
inclined sources have low $n$ values (i.e. disk-dominated), and large
$n$ values are more often less inclined.
Lastly, we compare the \gim total $m_{850}$ magnitudes with 
the model-independent determinations from
SExtractor (mag\_best).  We find a median offset of 0.29 mag (independent
of brightness) with a dispersion of 0.18 mag, in the sense that \gim
recovers total galaxy fluxes (by integrating the model profile to
infinity) that are systematically 
brighter than the SExtractor measurements. These
tests show that our fits, and the parameters that we derive from the fits,
are well-defined.

For each galaxy, \gim finds the best-fit (in a $\chi_{\nu}^2$ sense) \sersic 
model, produces model and residual images,
and converts each galaxy image into a set of structural
parameters including their internal measurement uncertainties.
In Figures \ref{Sthumbs} -- \ref{Lthumbs} 
we show GEMS F850LP-band galaxy and residual postage stamp
images for representative early-type examples selected randomly from
the three fixed-size bins out to $z\sim1$.
We note that the majority of galaxies have E/S0 visual morphologies,
and that an inspection of the fit residuals shows that the
the fitting does a fair job recovering the flux and overall profile shape.
The fitting results presented here ($r_{50}$ sizes and \sersic indices)
will be published in their entirety by H\"au{\ss}ler et al. (in preparation).

\bibliographystyle{/home/dmac/Papers/apj}
\bibliography{/home/dmac/Papers/references}

\clearpage


\onecolumn

\begin{deluxetable}{lcccl}
\tablewidth{0pt}
\tablenum{1}
\tabletypesize{\small}
\tablecolumns{5}
\tablecaption{Constrained Linear Fits to \lr and \mr Relations Evolution}
\tablehead{\colhead{Ordinate} & \multicolumn{3}{c}{Slopes ($m$)} & \colhead{Notes} \\
\cline{2-4} \\ 
\colhead{} & \colhead{{\sc i}} & \colhead{{\sc ii}} & \colhead{{\sc iii}} & \colhead{} \\
\colhead{(1)} & \colhead{(2)} & \colhead{(3)} & \colhead{(4)} & \colhead{(5)}}
\startdata
$\Delta M_V$   & $-1.22\pm 0.12$ & $-0.96\pm 0.06$ & $-0.65\pm 0.09$ & unfaded \\
               & $-1.65\pm 0.13$ & $-1.33\pm 0.06$ & $-0.72\pm 0.08$ & $z_{\rm form}=2$, [Fe/H] $=-0.1$\\
               & $-1.63\pm 0.12$ & $-1.28\pm 0.06$ & $-0.71\pm 0.08$ & $z_{\rm form}=3$, [Fe/H] $=-0.2$\\
               & $-1.59\pm 0.13$ & $-1.26\pm 0.06$ & $-0.71\pm 0.08$ & $z_{\rm form}=5$, [Fe/H] $=-0.3$\\
$\Delta \log_{10}{({\rm M}_{\star}h^2/{\rm M}_{\sun})}$ & $+0.24\pm 0.05$ & $+0.12\pm 0.03$ & $-0.13\pm 0.03$ & $g$-band-based MF\\
               & $+0.27\pm 0.05$ & $+0.14\pm 0.03$ & $-0.10\pm 0.03$ & $g_{\rm col}$-based MF\\
               & $+0.23\pm 0.05$ & $+0.10\pm 0.02$ & $-0.15\pm 0.03$ & $K$-band-based MF\\
\enddata
\tablecomments{We parameterize the amount of luminosity or stellar mass 
evolution with the least-squares slope from a best-fit linear relation to
the given ordinate (1) as a function of redshift, constrained to no
evolution at $z=0$.
For each galaxy size bin ({\sc i}, {\sc ii}, and {\sc iii}), we give the
best-fit slopes (2-4), including their $1\sigma$ dispersions 
from bootstrap resampling.  The number of galaxy measurements used in
each fit are $N_{\rm I}=102$, $N_{\rm II}=227$, and $N_{\rm III}=222$.
In (5) we list the different evolution analyses given by various
fadings of the selection function (for luminosity evolution), and by
various forms of the adopted stellar mass function (for stellar mass 
evolution).}
\label{LineFits}
\end{deluxetable}

\begin{deluxetable}{lcccccccl}
\tablewidth{0pt}
\tablenum{2}
\tabletypesize{\small}
\tablecolumns{9}
\tablecaption{Unconstrained Linear Fits to \lr and \mr Relations Evolution}
\tablehead{\colhead{Ordinate} & \multicolumn{3}{c}{Slopes ($m$)} & \colhead{} & \multicolumn{3}{c}{Intercepts ($b$)} & \colhead{Notes} \\
\cline{2-4} \cline{6-8} \\ 
\colhead{} & \colhead{{\sc i}} & \colhead{{\sc ii}} & \colhead{{\sc iii}} & \colhead{} & \colhead{{\sc i}} & \colhead{{\sc ii}} & \colhead{{\sc iii}} & \colhead{} \\
\colhead{(1)} & \colhead{(2)} & \colhead{(3)} & \colhead{(4)} & \colhead{} & \colhead{(5)} & \colhead{(6)} & \colhead{(7)} & \colhead{(8)}}
\startdata
$\Delta M_V$ & $-1.60\pm 0.46$ & $-0.68\pm 0.26$ & $-0.97\pm 0.27$ & & $+0.21\pm 0.21$ & $-0.18\pm 0.22$ & $+0.21\pm 0.16$ & unfaded\\
             & $-2.56\pm 0.56$ & $-1.65\pm 0.25$ & $-1.16\pm 0.34$ & & $+0.50\pm 0.28$ & $+0.20\pm 0.11$ & $+0.29\pm 0.25$ & $z_{\rm form} = 2$, [Fe/H] $=-0.1$\\
             & $-2.40\pm 0.54$ & $-1.47\pm 0.28$ & $-1.12\pm 0.30$ & & $+0.42\pm 0.28$ & $+0.13\pm 0.21$ & $+0.27\pm 0.16$ & $z_{\rm form} = 3$, [Fe/H] $=-0.2$\\
             & $-2.38\pm 0.52$ & $-1.39\pm 0.28$ & $-1.15\pm 0.32$ & & $+0.43\pm 0.26$ & $+0.08\pm 0.18$ & $+0.29\pm 0.16$ & $z_{\rm form} = 5$, [Fe/H] $=-0.3$\\
$\Delta \log_{10}{({\rm M}_{\star}h^2/{\rm M}_{\sun})}$ & $+0.69\pm 0.20$ & $+0.35\pm 0.11$ & $+0.16\pm 0.12$ & & $-0.24\pm 0.11$ & $-0.15\pm 0.06$ & $-0.20\pm 0.06$ & $g$-band-based MF\\
                  & $+0.63\pm 0.22$ & $+0.35\pm 0.13$ & $+0.19\pm 0.12$ & & $-0.20\pm 0.14$ & $-0.14\pm 0.07$ & $-0.19\pm 0.07$ & $g_{\rm col}$-based MF\\
                  & $+0.69\pm 0.22$ & $+0.37\pm 0.13$ & $+0.17\pm 0.12$ & & $-0.25\pm 0.12$ & $-0.18\pm 0.08$ & $-0.21\pm 0.05$ & $K$-band-based MF\\
\enddata
\tablecomments{The unconstrained linear fit (least-squares) parameters
for luminosity and stellar mass evolution.  
For each ordinate (1) as a function of redshift, we give the
best-fit slopes (2-4) and intercepts (5-7), including their $1\sigma$
dispersions from bootstrap resampling. As in Table \ref{LineFits},
we give the parameters for each galaxy size bin
({\sc i}, {\sc ii}, and {\sc iii}) and each evolution analyses (5). For
the unconstrained case we used
the same number of galaxy measurements as with the constrained case.
}
\label{LineFits_un}
\end{deluxetable}

\end{document}